\documentclass[useAMS,usenatbib]{mn2e}
\usepackage{graphicx}
\usepackage{color}
\usepackage{amsmath}

\title[A treatment procedure for NIFS data cubes]{A treatment procedure for Gemini North/NIFS data cubes: application to NGC 4151}

\author[Menezes et al.] {R.~B.~Menezes,$^1$\thanks{E-mail:
robertobm@astro.iag.usp.br} J.~E.~Steiner$^1$ and T.~V.~Ricci$^1$ \\
$^{1}$Instituto de Astronomia Geof\'isica e Ci\^encias Atmosf\'ericas, Universidade de S\~ao Paulo, Rua do Mat\~ao 1226, \\Cidade Universit\'aria, S\~ao Paulo, SP CEP 05508-090, Brazil}

\begin{document}

\date{Accepted 2013 December 7. Received 2013 December 7; in original form 2013 July 11}

\pagerange{\pageref{firstpage}--\pageref{lastpage}} \pubyear{2014}

\maketitle

\label{firstpage}

\begin{abstract}

We present a detailed procedure for treating data cubes obtained with the Near-Infrared Integral Field Spectrograph (NIFS) of the Gemini North telescope. This process includes the following steps: correction of the differential atmospheric refraction, spatial re-sampling, Butterworth spatial filtering, `instrumental fingerprint' removal and Richardson-Lucy deconvolution. The clearer contours of the structures obtained with the spatial re-sampling, the high spatial-frequency noise removed with the Butterworth spatial filtering, the removed `instrumental fingerprints' (which take the form of vertical stripes along the images) and the improvement of the spatial resolution obtained with the Richardson-Lucy deconvolution result in images with a considerably higher quality. An image of the Br$\gamma$ emission line from the treated data cube of NGC 4151 allows the detection of individual ionized-gas clouds (almost undetectable without the treatment procedure) of the narrow-line region of this galaxy, which are also seen in an [O III] image obtained with the \textit{Hubble Space Telescope}. The radial velocities determined for each one of these clouds seem to be compatible with models of biconical outflows proposed by previous studies. Considering the observed improvements, we believe that the procedure we describe in this work may result in more reliable analysis of data obtained with this instrument.

\end{abstract}


\section{Introduction}

\begin{figure*}
\begin{center}
  \includegraphics[scale=0.40]{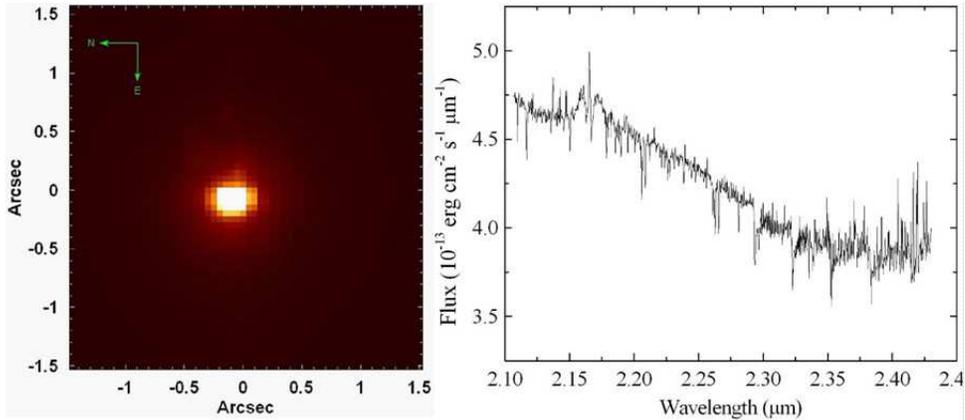}
  \caption{Left: image of one of the nine data cubes of GQ Lup collapsed along the spectral axis. Right: average spectrum of the data cube shown at left.\label{fig1}}
\end{center}
\end{figure*}

Data cubes are large data sets with two spatial dimensions and one spectral dimension. As a consequence, in a data cube, one can obtain images of an object at different wavelengths or spectra of different spatial regions of this object. With the advent of instruments like Integral Field Spectrographs (IFSs) and modern Fabry-Perot spectrographs in the last decades, data cubes have been produced with increasing pace. Due to the high dimensions of these data sets, it is usually hard to fully analyse them using traditional methods and, therefore, only a subset of the data ends up being taken into account (kinematical maps, line flux maps, etc). We have developed PCA Tomography \citep{ste09}, a procedure that consists of applying principal component analysis (PCA; Murtagh \& Heck 1987; Fukunaga 1990) to data cubes, in order to extract information from them in an optimized way. However, a powerful analysis tool on its own is not enough to work with data cubes. These data sets are usually contaminated by a series of artefacts (like high spatial-frequency noise, for example), often related to the instruments involved, which make it difficult to perform a detailed analysis. Therefore, in order to extract reliable information from data cubes, it is advisable to remove these artefacts with a proper data treatment.

The Near-Infrared Integral Field Spectrograph (NIFS; McGregor et al. 2002) of the Gemini North telescope provides 3D imaging spectroscopy in a spectral range of 0.95 to 2.4 $\mu m$ (covering the spectral bands \textit{Z}, \textit{J}, \textit{H} and \textit{K}), with a field of view (FOV) of 3.0 arcsec $\times$ 3.0 arcsec. NIFS is designed to work in association with the adaptive optics (AO) system Altair, resulting in observations with spatial resolutions of about 0.1 arcsec. The incident light in the FOV of NIFS passes to an image slicer, where 29 small concave mirrors direct the light to another array of concave mirrors, called pupil mirrors. This second group of mirrors redirects all the light to a third array of mirrors, called field mirrors. However, besides redirecting the light, the pupil mirrors also reconfigure it in such a way that, when the light reaches the field mirrors, it is disposed as a long thin slit. With this new configuration, the light passes through a collimator, through a collimator corrector, through the grating and finally, it is directed to the detector. All of the optical and geometric properties of NIFS result in rectangular spatial pixels of 0.103 arcsec $\times$ 0.043 arcsec.

This is the first of a series of papers in which we will describe data treatment techniques, developed by us \citep{men12}, to be applied to data cubes obtained with different instruments. In this paper, we present a detailed procedure for treating NIFS data cubes, after the data reduction. This procedure is presented in a specific sequence, which, according to our tests, gives optimized results. The main purpose of this data treatment is to remove artefacts typically seen in NIFS data cubes, which can affect considerably the data analysis. In Section 2, we describe the observations and the reduction of the data used in this paper to describe the procedure. In Section 3, we discuss the correction of the differential atmospheric refraction (DAR). In Section 4, we describe the spatial re-sampling of the data. In Section 5, we present the Butterworth spatial filtering, used to remove high spatial-frequency noise. In Section 6, we describe the process for removing `instrumental fingerprints'. In Section 7, we discuss the Richardson-Lucy deconvolution, used to improve the spatial resolution of the observations. In Section 8, we present the analysis of the data cubes of NGC 4151 as a scientific example illustrating the benefits obtained with the data treatment methodology we describe. Finally, we draw a summary and our conclusions in Section 9.

\section{Observations and data reduction}

The data used in most of this paper to show the data treatment procedure are observations of the star GQ Lup, available in the NIFS public access data bank (\textit{www3.cadc-ccda.hia-iha.nrc-cnrc.gc.ca/gsa/}). The observations were made in the \textit{K} band on 2007 June 27, (Programme: GN-2007A-Q-46; PI: J-F. Lavigne), using a natural guide star (in this case, GQ Lup) for the AO. A total of nine 50 s exposures of GQ Lup were retrieved from the data bank to be used in this work. In order to analyse the spatial displacements along the spectral axis of NIFS data cubes, we also retrieved observations of this star in the \textit{J} and \textit{H} bands. One 400 s exposure in the \textit{J} band and another 300 s exposure in the \textit{H} band of GQ Lup (both from the same programme mentioned before and using the observed star for the guiding of the AO) were retrieved. The observation in the \textit{J} band was made on 2007 May 30, and the observation in the \textit{H} band was made on 2007 June 1. In order to show the process for removing `instrumental fingerprints' (in Section 6), we used observations of the nuclear region of the galaxy M87. These observations were made in the \textit{K} band on 2008 April 22 (Programme: GN-2008A-Q-12; PI: D. Richstone). A total of nine 600 s exposures of the nuclear region of M87 were obtained. Finally, in order to show a scientific example that illustrates the benefits obtained with the data treatment methodology we describe, we used observations of the nuclear region of the galaxy NGC 4151. These observations were made in the \textit{K} band on 2006 December 12 (Programme: GN-2006B-C-9; PI: P. McGregor). A total of eight 90 s exposures were obtained. Although we show the data treatment process in this paper using, mainly, data in the \textit{K} band, the procedure is entirely analogous for data in the \textit{Z}, \textit{J} and \textit{H} bands.  

\begin{figure*}
\begin{center}
  \includegraphics[scale=0.19]{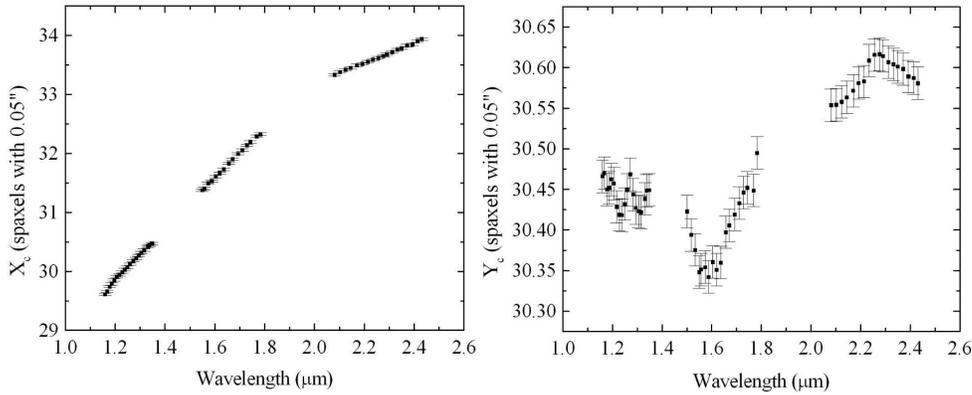}
  \caption{Graphs of the coordinates $X_c$ and $Y_c$ of the centres of the star GQ Lup in three data cubes, obtained after the data reduction, in the \textit{J}, \textit{H} and \textit{K} bands (one data cube for each band). There is no spatial dithering between the data cubes.\label{fig2}}
\end{center}
\end{figure*}

Standard calibration images of flat-field, dark flat-field (used in the data reduction for the subtraction of the dark current), Ronchi-flat (used in the data reduction to compute spatial distortions and to perform a spatial rectification), arc lamp and sky field were obtained to be used in the data reduction. Calibration images of A0V standard stars were also used. The data reduction was made in the IRAF environment and included the following steps: determination of the trim, sky subtraction, bad pixel correction, flat-field correction, spatial rectification, wavelength calibration, telluric absorption removal, flux calibration (assuming that the used A0V standard star has a blackbody spectrum) and data cube construction. At the end of this process, we obtained 11 data cubes of GQ Lup (9 in the \textit{K} band, 1 in the \textit{J} band and one in the \textit{H} band), 9 data cubes of M87 in the \textit{K} band and 8 data cubes of NGC 4151 in the \textit{K} band. All of these data cubes were reduced with spatial pixels (spaxels) of 0.05 arcsec $\times$ 0.05 arcsec. Fig.~\ref{fig1} shows an image of one of the nine data cubes of GQ Lup in the \textit{K} band collapsed along the spectral axis and also its average spectrum. This data cube actually extends to 2.0 $\mu m$ at the blue. However, due to some flux calibration problems in that part of the spectrum, we decided to not discuss it in this paper, except in Section 2, where we included this blue region of the spectrum to analyse the spatial displacements of the object along the spectral axis of the data cube.

\section{Correction of the DAR}

It is known that the refraction caused by the Earth's atmosphere affects the light emitted by any celestial object that is observed by ground-based telescopes. The atmospheric refraction alters the original trajectory of the light and, as a consequence, an observed object looks closer to the zenith than it really is. This effect depends on the wavelength of the light and can affect considerably astronomical observations made from the ground.

Mathematically, we can say that the main consequence of the atmospheric refraction for the observation of any object is that the original zenith distance $z$ of the light incident at the top of the atmosphere is higher than the observed zenith distance $\zeta$. The angle $R = z - \zeta$ is called refraction angle and is given by the following expression \citep{sma31}:

\begin{equation}
R = 206265 \times \left(\mu_0 - 1\right) \times tan \zeta,
\end{equation}
where $\mu_0$ is the refraction index close to the Earth's surface.

Assuming an atmospheric pressure $P = 10^5$ Pa, a temperature $T = 20 \degr C$, and a CO$_2$ volume fraction of 0.0004, with low humidity, the refraction index $\mu_0$ can be given by \citep{bon98}

\begin{equation}
\resizebox{.93\hsize}{!}{$\left(\mu\left(\lambda\right)_{20,10^5} - 1\right) \times 10^8 = 8091.37 + \frac{2333983}{130 - \left(\frac{1}{\lambda}\right)^2}+\frac{15518}{38.9 - \left(\frac{1}{\lambda}\right)^2}$},
\end{equation}
where $\lambda$ is the wavelength (in microns).

However, most of the ground-based observatories are located at high altitudes; therefore, the proper refraction indexes for these locations must be calculated for lower values of temperature and pressure. Under these conditions (keeping the CO$_2$ volume fraction in 0.0004), the refraction index is given by \citep{bon98}

\begin{equation}
\resizebox{7.8 cm}{0.4 cm}{$\left(\mu\left(\lambda\right)_{T,P}-1\right) = \left(\mu\left(\lambda\right)_{20,10^5}-1\right) \times \frac{P \times \left[1+\left(0.5953-0.009876 \times T\right) \times 10^{-8} \times P\right]}{93214.60 \times \left(1 + 0.003661 \times T\right)}$},
\end{equation}
where $P$ is measured in Pa and $T$ is measured in $\degr C$.

In the presence of water vapour, the value $\mu(\lambda)_{T,P}$ is reduced by a factor \citep{bon98}

\begin{equation}
f \times \left(3.8020 - \frac{0.0384}{\lambda^2}\right) \times 10^{-10},
\end{equation}
where $f$ is the water vapour pressure measured in Pa and $\lambda$ is measured in microns.

Considering the dependence of $\mu_0$ on wavelength, the change suffered by the refraction angle when the wavelength varies from $\lambda_1$ to $\lambda_2$ (keeping $\zeta$ constant) can be given by

\begin{equation}
\Delta R = R\left(\lambda_2\right) - R\left(\lambda_1\right) 
\end{equation}

\begin{equation}
\Delta R = 206265 \times \left[\mu_0\left(\lambda_2\right) - \mu_0\left(\lambda_1\right)\right] \times tan \zeta
\end{equation}

Usually, during the observation of a given object, one can assume that the zenith distance and other parameters, like pressure and temperature, are, approximately, constant (as long as the exposure time is not very high). However, the wavelength is a variable parameter in a data cube, as the image of the observed object can be seen in different spectral regions. Due to the dependence of $R$ on wavelength, one can conclude that the refraction angle varies along the spectral axis of the data cube, the origin of the so-called DAR. The main characteristic of this effect is that it changes the position of the observed object along the spectral axis of the data cube.

The DAR is considerably higher in the optical than in the infrared (see equation 2). However, as can be seen in the previous equations, other parameters (zenith distance, temperature, atmospheric pressure and water vapour pressure), besides the wavelength, have an important influence in this effect. The DAR is also higher, for example, for objects at a higher zenith distance (see equation 1). For more details about the DAR effect, see \citet{fil82}.

\begin{figure}
\begin{center}
  \includegraphics[scale=0.30]{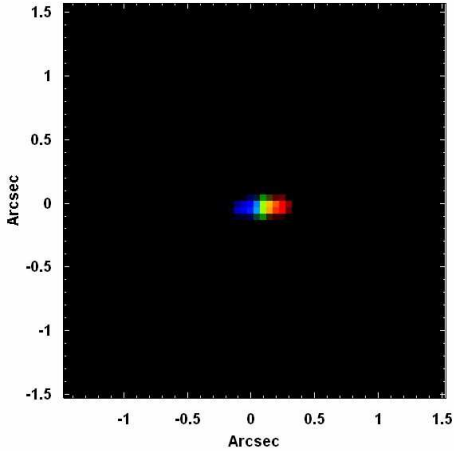}
  \caption{RGB composition containing images of GQ Lup in the \textit{J}, \textit{H} and \textit{K} bands. The blue colour represents the wavelength range 1.1600 - 1.1615 $\mu m$, the green colour represents the wavelength range 1.6550 - 1.6565 $\mu m$, and the red colour represents the wavelength range 2.4250 - 2.4265 $\mu m$.\label{fig3}}
\end{center}
\end{figure}

After the data reduction process, the first step in the proposed sequence for treating NIFS data cubes is the correction of the DAR. The main purpose of this section, however, is not to present a new method for calculating the DAR effect, as, in our data treatment procedure, we simply apply, when it is necessary, the equations from \citet{bon98} and \citet{fil82} to perform this calculation. The main purposes of this section are the following: to show that, although the DAR is lower in the infrared than in the optical, the spatial displacements of the observed objects along the spectral axis of NIFS data cubes may be significant for some studies; to analyse, in detail, the spatial displacements along the spectral axis of NIFS data cubes, in order to evaluate if the behaviour of these displacements is compatible with the DAR effect.

Fig.~\ref{fig2} shows the graphs of the coordinates $X_c$ and $Y_c$ of the centres of the star GQ Lup in three data cubes, obtained after the data reduction, in the \textit{J}, \textit{H} and \textit{K} bands (one data cube for each band). There is no spatial dithering between these three data cubes. We can see that the total displacement of the centre of the star from the \textit{J} band to the \textit{K} band is $\sim$ 0.2 arcsec along the horizontal axis and $\sim$ 0.015 arcsec along the vertical axis. Since the total displacement along the horizontal axis is twice the typical NIFS spatial resolution (0.1 arcsec), we conclude that such displacement is significant for studies involving the entire continuum of the \textit{J}, \textit{H} and \textit{K} bands. One example of such a study is a spectral synthesis, which is usually performed using data in the \textit{J}, \textit{H} and \textit{K} bands together. Studies involving spectral lines may also be affected by this effect if one wishes to compare images of lines located in different spectral bands. One example would be studies of emission line ratios. Fig.~\ref{fig3} shows an RGB composition containing images of GQ Lup in the \textit{J} (blue), \textit{H} (green) and \textit{K} (red) bands. This RGB makes it easier to see the scientific necessity of the correction of the DAR effect, when the spectral continua of the \textit{J}, \textit{H} and \textit{K} bands are considered simultaneously. On the other hand, in Fig.~\ref{fig2}, we can see that the horizontal displacements in the \textit{J}, \textit{H} and \textit{K} bands are $\sim$ 0.04, $\sim$ 0.05 and $\sim$ 0.03 arcsec, respectively, and the vertical displacements in these bands are $\sim$ 0.003, $\sim$ 0.008 and $\sim$ 0.003 arcsec, respectively. These spatial displacements are significantly smaller than the typical NIFS spatial resolution, indicating that studies involving only one of these spectral bands may not require a correction of this effect (although a horizontal displacement of $\sim$ 0.05 arcsec, for example, which is equal to 50 per cent of the typical NIFS spatial resolution, may be significant when accurate astrometry is required). Considering all of that, we conclude that the correction of the DAR effect in NIFS data cubes may be important in certain cases, especially for studies involving the entire spectral continuum in the \textit{J}, \textit{H} and \textit{K} bands. It is also important to mention that the zenith distances of the data cubes of GQ Lup in the \textit{J}, \textit{H} and \textit{K} bands are $55.42\degr$, $55.38\degr$, and $56.55\degr$, respectively, which are relatively high values for a zenith distance. Observations with lower zenith distances will have smaller spatial displacements along the spectral axis of the data cubes, and in these cases, even studies involving the spectral continua in the \textit{J}, \textit{H} and \textit{K} bands simultaneously may not require a correction of the DAR effect. The fact that the spatial displacements along the spectral axis of NIFS data cubes are significant in certain cases may be a little surprising, as a correction of the DAR effect is usually ignored in the infrared. However, this can be explained by the fact that NIFS observations are usually performed using the AO from Altair, resulting in spatial resolutions of $\sim$ 0.1 arcsec. Therefore, although the DAR is small in the infrared, the high spatial resolutions of NIFS observations provided by the AO make this effect significant in certain cases.

\begin{figure*}
\begin{center}
  \includegraphics[scale=0.41]{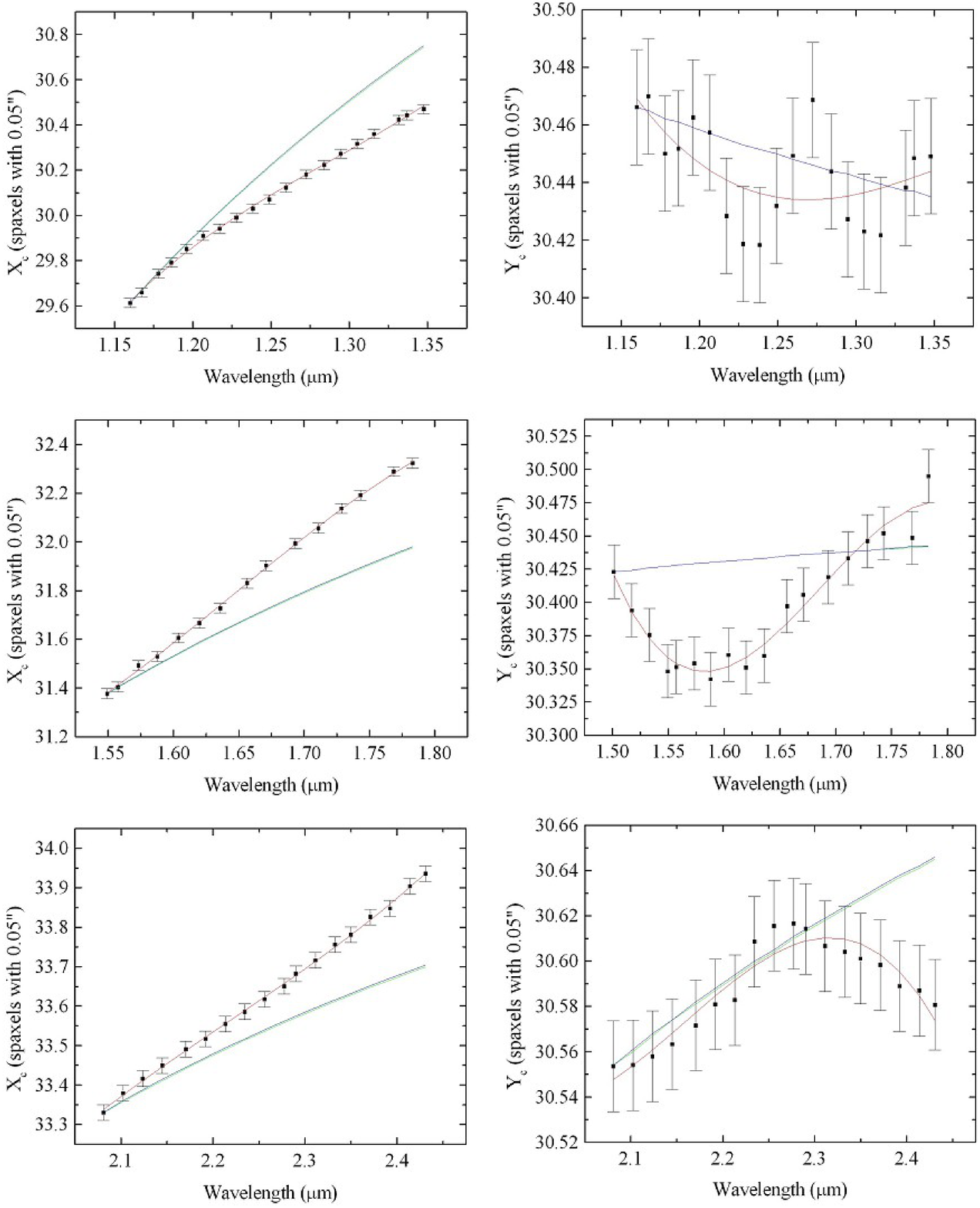}
  \caption{Graphs of the coordinates $X_c$ and $Y_c$ of the centres of the star GQ Lup in three data cubes, obtained after the data reduction, in the \textit{J}, \textit{H} and \textit{K} bands (one data cube for each band). The third degree polynomials fitted to the points are shown in red, the theoretical curves obtained assuming a plane-parallel atmosphere are shown in green and the theoretical curves obtained with the task `\textit{refro}' from the `SLALIB' package are shown in blue.\label{fig4}}
\end{center}
\end{figure*}

\begin{figure*}
\begin{center}
  \includegraphics[scale=0.41]{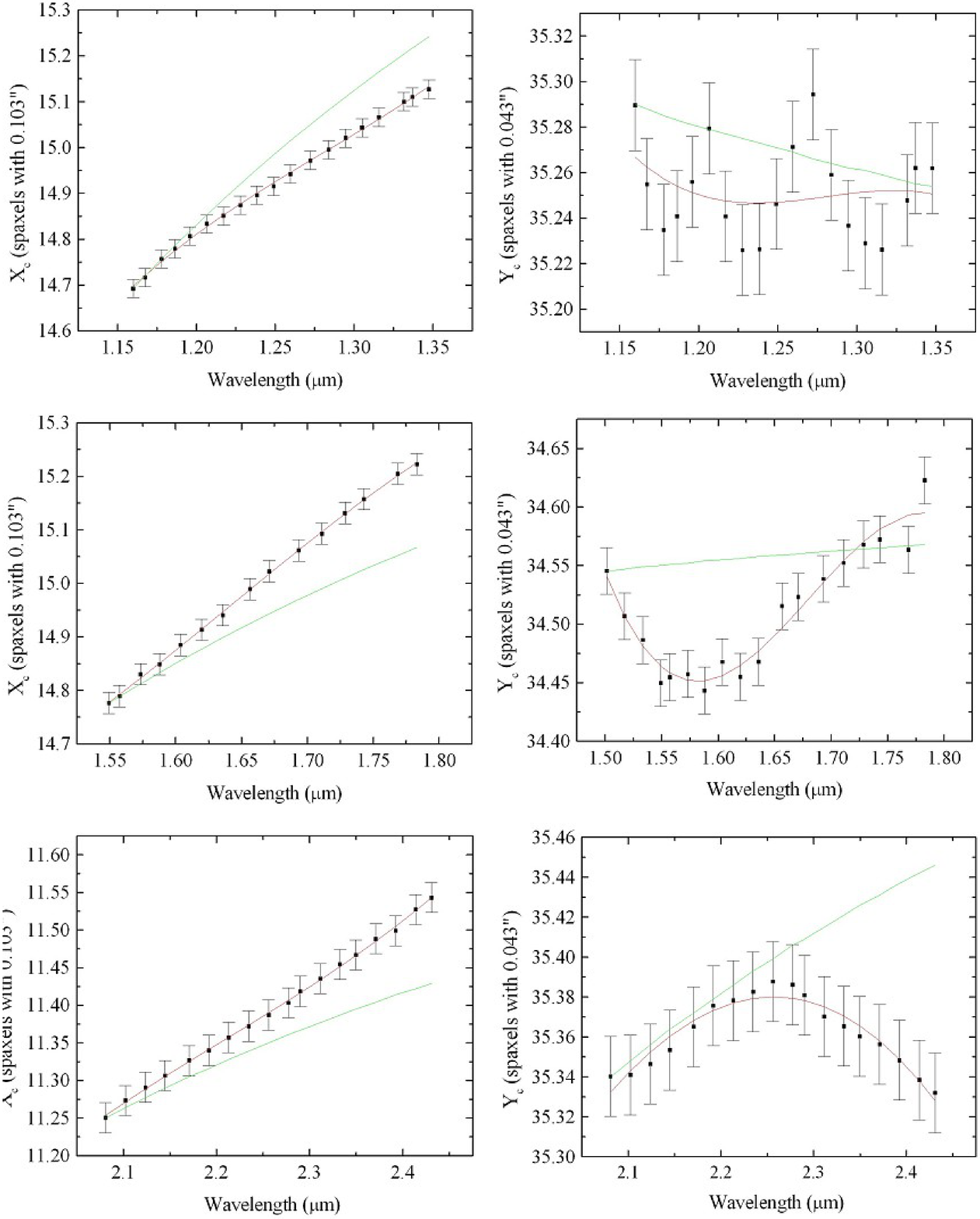}
  \caption{Graphs of the coordinates $X_c$ and $Y_c$ of the centres of the star GQ Lup in three data cubes, obtained after the data reduction, in the \textit{J}, \textit{H} and \textit{K} bands (one data cube for each band), keeping the original size (0.103 arcsec $\times$ 0.043 arcsec) of their spaxels. The third degree polynomials fitted to the points are shown in red and the theoretical curves obtained assuming a plane-parallel atmosphere are shown in green.\label{fig5}}
\end{center}
\end{figure*}

\begin{figure*}
\begin{center}
  \includegraphics[scale=0.41]{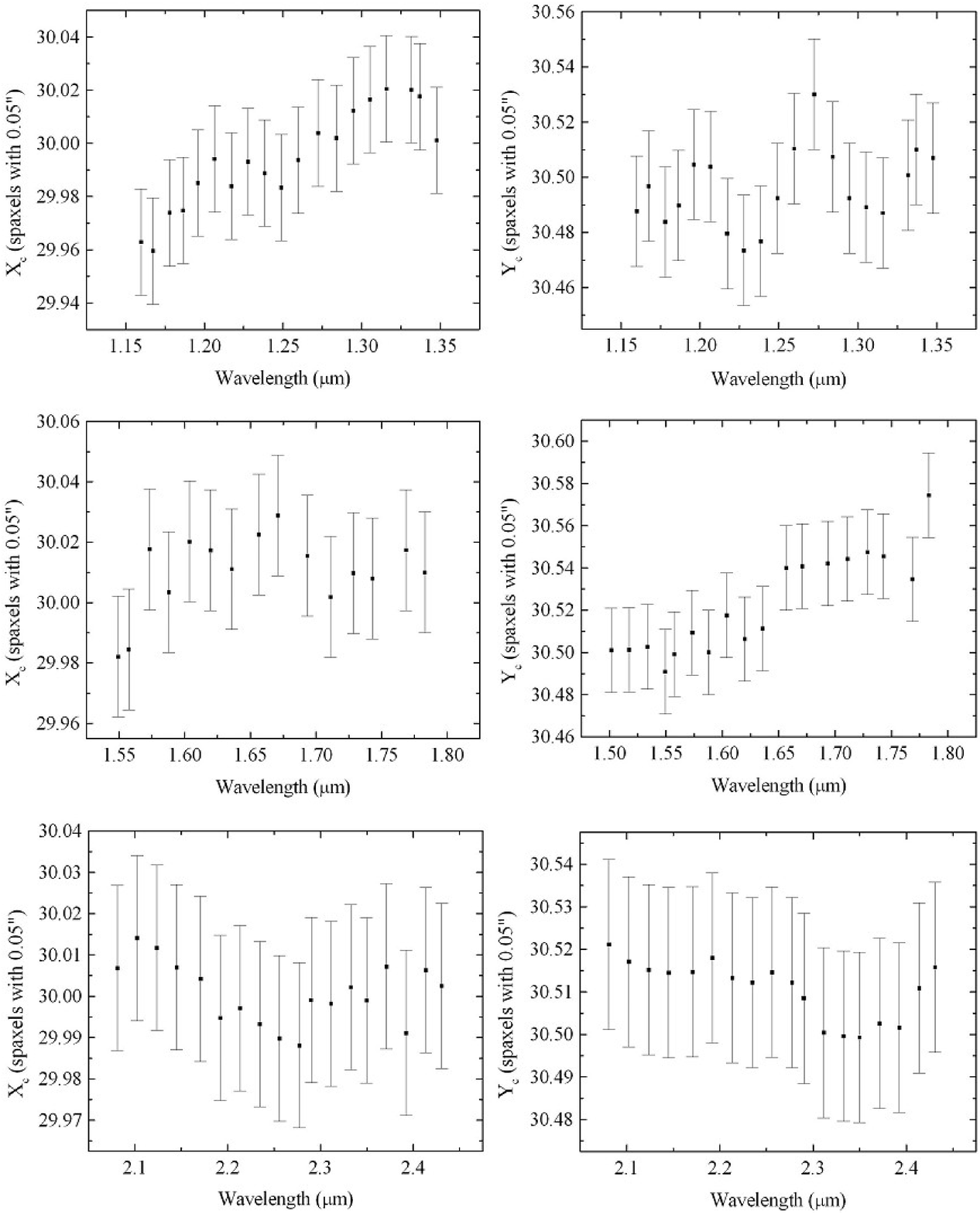}
  \caption{Graphs of the coordinates $X_c$ and $Y_c$ of the centres of the star GQ Lup in three data cubes, obtained after the data reduction, in the \textit{J}, \textit{H} and \textit{K} bands (one data cube for each band), after the correction of the spatial displacements along the spectral axis, using the practical approach.\label{fig6}}
\end{center}
\end{figure*}

With the next generation of extremely large telescopes, the AO-corrected point-spread functions (PSFs) obtained in the observations will be significantly smaller. As a consequence, the DAR will have a larger impact on data cubes, even in situations where this effect is usually ignored, like observations in the infrared, for example. This fact certainly must be addressed in the data reduction procedures for these future data sets.

In order to perform the correction of the DAR effect, we use a script, written in Interactive Data Language (IDL), based on an algorithm developed by us. This script shifts each image of the data cube in a way that, at the end of the process, the position of the observed object does not change along the spectral axis of the data cube. The correction of the DAR can be done using two strategies: the theoretical approach and the practical approach. In the theoretical approach, the values of $\Delta R$ (along the spectral axis) of the centres of a given structure in the data cube are calculated using the theoretical equations of the DAR effect. After that, in order to obtain the coordinates $X_c$ and $Y_c$ of the centres of the structure, the values of $\Delta R$ are projected on the horizontal and vertical axis of the FOV, using the parallactic angle $\eta$ (which is given by the equations in Filippenko 1982). Using the simulated coordinates, the algorithm calculates and applies shifts to the images of the data cube, in order to remove the spatial displacement of the structure along the spectral axis. On the other hand, in the practical approach, the coordinates of the centres of a given structure are not calculated, but are measured directly. The practical approach is usually performed in the following way: first, wavelength intervals along the spectral axis of the data cube are chosen. Usually, 20 or 25 intervals are enough and it is advisable that these intervals do not contain spectral lines. The images in these intervals are, then, summed (resulting in one image for each interval) and the coordinates of the centres of a given structure in each one of the obtained images are determined using a proper method (one possibility is the task `\textit{imexamine}' of the IRAF software). After that, the coordinates $X_c$ and $Y_c$ of the centres are plotted as a function of the wavelength (it is important to mention that, since each image results from the sum of many others in each one of the wavelength intervals, the elaborated graphs associate the coordinates $X_c$ and $Y_c$ to the mean wavelengths corresponding to each image). Third degree polynomials are fitted to the graphs of $X_c \times \lambda$ and $Y_c \times \lambda$. Finally, using the coefficients of the third degree polynomials obtained, the script calculates and applies shifts to the images of the data cube, in order to remove the spatial displacement of the structure along the spectral axis. It is important to emphasize that our script does not use any new method for calculating the DAR effect (only the equations from B\"onsch \& Potulski 1998 and Filippenko 1982). The main advantages of our algorithm are that it can be used with both the theoretical and the practical approaches and that, at the end of the procedure, the position of the observed object remains stable along the spectral axis of an NIFS data cube with a precision higher than 0.01 arcsec (see more details below). However, there are many other algorithms capable of shifting each image of a data cube, in order to remove spatial displacements along the spectral axis. Many of these algorithms (like the task `\textit{imshift}' of the IRAF software) are essentially as precise as our algorithm and can be used in this treatment procedure.

The theoretical approach is advisable when the theoretical equations of the DAR effect reproduce the exact behaviour of the spatial displacement of a given structure along the spectral axis of the data cube. However, if (for any reason) that is not the case, the practical approach should be used.

Equation (1) was obtained assuming a plane-parallel atmosphere. However, there are more precise routines that perform this calculation taking into account the curvature of the Earth's atmosphere. One example of these routines is the task `\textit{refro}' from the `SLALIB' package. Fig.~\ref{fig4} shows the graphs of $X_c \times \lambda$ and $Y_c \times \lambda$ of three data cubes of GQ Lup, obtained after the data reduction, in the \textit{J}, \textit{H}, and \textit{K} bands (one data cube for each band). In these graphs, we can also see the third degree polynomials fitted to the points, the theoretical curves obtained assuming a plane-parallel atmosphere (using the equations from B\"onsch \& Potulski 1998 and Filippenko 1982), and the theoretical curves obtained with the task `\textit{refro}' (since the task `\textit{refro}' only calculates the values of $\Delta R$, we projected them on the horizontal and vertical axis of the FOV, using the parallactic angle $\eta$, which is given by the equations in Filippenko 1982).

The first thing we can notice in Fig.~\ref{fig4} is that the theoretical curves obtained assuming a plane-parallel atmosphere and taking into account the curvature of the atmosphere were very similar to each other. In some graphs, we cannot even differentiate these two curves. Besides that, none of the theoretical curves reproduce properly the values of $X_c$ along the spectral axis of the data cubes. There are also discrepancies between the theoretical curves and the values of $Y_c$ along the spectral axis of the data cubes in the \textit{H} and \textit{K} bands. In the case of the data cube in the \textit{J} band, although the theoretical curves do not reproduce the exact behaviour of the points in the graph of $Y_c \times \lambda$, there is no clear discrepancy, as the curves are within the error bars of the points. Our previous experiences revealed that the discrepancies observed in Fig.~\ref{fig4} are very common in NIFS data cubes. Considering all of that, we conclude that the theoretical curves of the DAR effect obtained assuming a plane-parallel atmosphere or taking into account the curvature of the atmosphere do not reproduce the exact behaviour of the spatial displacements of structures along the spectral axis of NIFS data cubes.

One natural question at this point is: what is the cause of the discrepancies observed in Fig.~\ref{fig4} ? Clearly, they cannot be attributed to a simplistic calculation of the DAR effect, as even the task `\textit{refro}' (which performs a very detailed calculation of the DAR effect, considering both the curvature of the Earth's atmosphere and an approximation to the change in atmospheric temperature and pressure with altitude) could not reproduce the observed behaviour of the spatial displacements. One possibility is that the discrepancies were caused by the fact that, during the data reduction, there was a spatial re-sampling to spaxels of 0.05 arcsec $\times$ 0.05 arcsec, from the 0.103 arcsec $\times$ 0.043 arcsec raw spaxels. In order to evaluate this hypothesis, we reduced again the data cubes in the \textit{J}, \textit{H} and \textit{K} bands (with the IRAF software), but, this time, keeping the original size of the spaxels (0.103 arcsec $\times$ 0.043 arcsec). After that, we determined the values of $X_c$ and $Y_c$ along the spectral axis of the data cubes. The results are shown in Fig.~\ref{fig5}. We can see that the discrepancies observed in Fig.~\ref{fig4} between the coordinates of the centres of GQ Lup and the theoretical curves obtained assuming a plane-parallel atmosphere (which are essentially equal to the theoretical curves obtained taking into account the curvature of the Earth's atmosphere) remain practically unchanged.

Considering the results obtained in Fig.~\ref{fig4} and in Fig.~\ref{fig5}, we conclude that the discrepancies between the coordinates of the centres of GQ Lup and the theoretical curves of the DAR effect cannot be attributed to a simplistic calculation of the effect or to the spatial re-sampling to spaxels of 0.05 arcsec $\times$ 0.05 arcsec during the data reduction. The exact cause of these discrepancies remains unknown. One possibility is that this is an instrumental effect. If this hypothesis is correct, the spatial displacements of the structures along the spectral axis of NIFS data cubes may even depend on observational parameters and an analysis of this effect would require the use of data cubes from different observations. Such analysis is beyond the scope of this paper and will not be performed here.

Since the theoretical curves do not reproduce properly the spatial displacements along the spectral axis, the practical approach is the most precise for removing this effect from NIFS data cubes. However, as Fig.~\ref{fig4} shows, the discrepancies between the coordinates of the centres of GQ Lup and the theoretical curves are all lower than 0.02 arcsec. In our previous experiences, we verified that it is very difficult to observe discrepancies higher than this. Since this value is significantly lower than the typical NIFS spatial resolution, we conclude that the theoretical approach may also be used to remove, with a good precision, the spatial displacements of structures along the spectral axis of NIFS data cubes. 

Using the practical approach, we applied our algorithm in order to remove the spatial displacements along the spectral axis of the three data cubes of GQ Lup used to generate Fig.~\ref{fig4} and Fig.~\ref{fig5}. After that, we measured again the coordinates $X_c$ and $Y_c$ of the star along the spectral axis of these three data cubes. The results are shown in Fig.~\ref{fig6}. We can see that the procedure reduced considerably the total displacement of $X_c$ along the spectral axis of the data cubes. In the case of $Y_c$, we did not expect considerable improvements, since the maximum displacement observed in this coordinate in Fig.~\ref{fig4} and Fig.~\ref{fig5} was close to 0.008 arcsec and the precision of our algorithm is usually not higher than 0.005 arcsec. Nevertheless, Fig.~\ref{fig6} shows that the procedure also reduced the total displacement of $Y_c$ along the spectral axis of the data cubes, although the gain was very subtle. The expected values, after the correction, for $X_c$ and $Y_c$ were 30.0 and 30.5, respectively. Therefore, we conclude that the precision ($1\sigma$) of our algorithm, in this case, was higher than 0.001 arcsec in $X_c$ and higher than 0.002 arcsec in $Y_c$. Usually, the algorithm results in precisions between 0.001 arcsec and 0.01 arcsec, for NIFS data cubes.

Using the practical approach, we also applied our algorithm in order to remove the spatial displacements along the spectral axis of the other data cubes of GQ Lup in the \textit{K} band. After that, in order to combine the nine data cubes in the \textit{K} band into one, they were divided in three groups with three data cubes each. A median of the data cubes of each group was calculated, resulting in three data cubes at the end of the process. Finally, we calculated the median of these three data cubes and obtained the combined data cube.

\section{Spatial re-sampling}

\begin{figure*}
\begin{center}
  \includegraphics[scale=0.42]{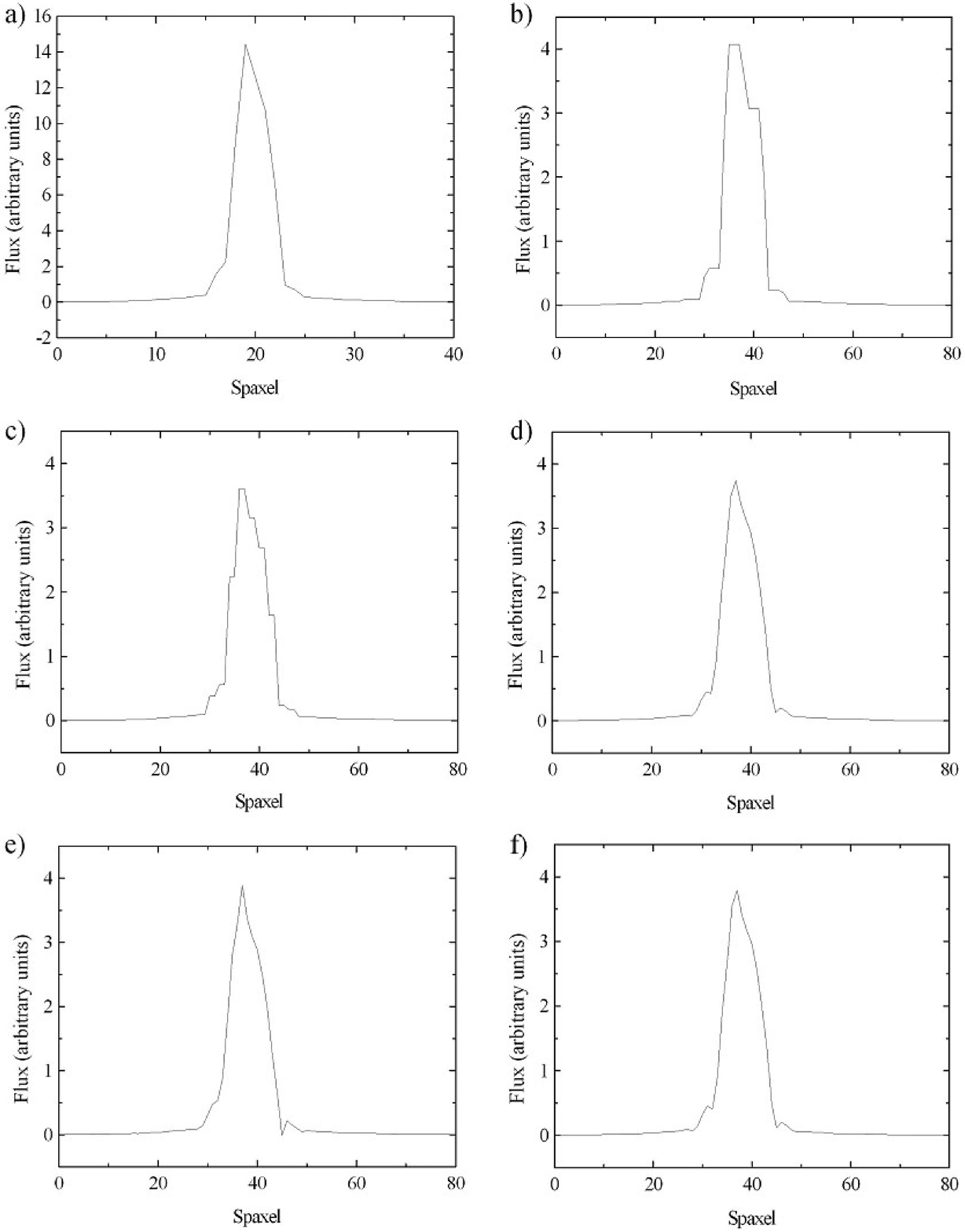}
  \caption{Graphs of the horizontal brightness profiles of the images of a collapsed intermediate-wavelength interval of the data cube of GQ Lup a) with spaxels of 0.05 arcsec $\times$ 0.05 arcsec without the spatial re-sampling, b) with spaxels of 0.025 arcsec $\times$ 0.025 arcsec obtained with a different data reduction procedure (which originated the spaxels of 0.025 arcsec $\times$ 0.025 arcsec), c) with spaxels of 0.025 arcsec $\times$ 0.025 arcsec obtained with a simple spatial re-sampling without any interpolation, d) with spaxels of 0.025 arcsec $\times$ 0.025 arcsec obtained with a spatial re-sampling followed by a lsquadratic interpolation, e) with spaxels of 0.025 arcsec $\times$ 0.025 arcsec obtained with a spatial re-sampling followed by a quadratic interpolation and f) with spaxels of 0.025 arcsec $\times$ 0.025 arcsec obtained with a spatial re-sampling followed by a spline interpolation.\label{fig7}}
\end{center}
\end{figure*}

\begin{figure*}
\begin{center}
  \includegraphics[scale=0.45]{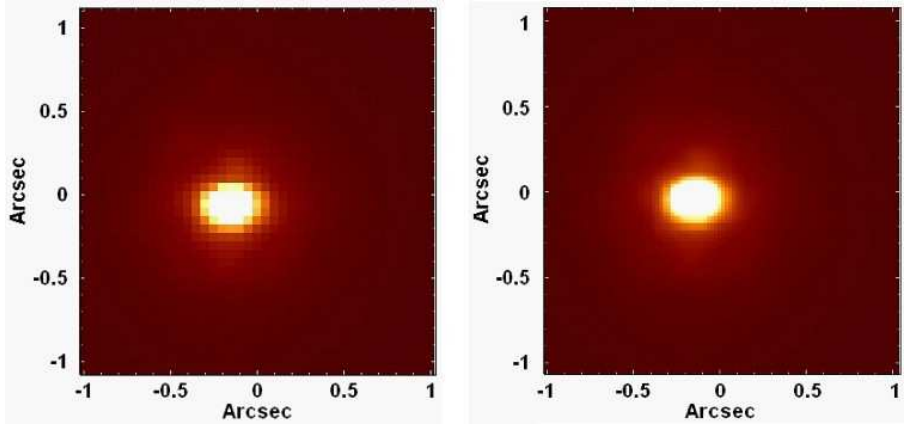}
  \caption{Left: image of a collapsed intermediate-wavelength interval of the data cube of GQ Lup, before the spatial re-sampling. Right: image of the same collapsed wavelength interval of the image at left, after the spatial re-sampling.\label{fig8}}
\end{center}
\end{figure*}

\begin{figure*}
\begin{center}
  \includegraphics[scale=0.45]{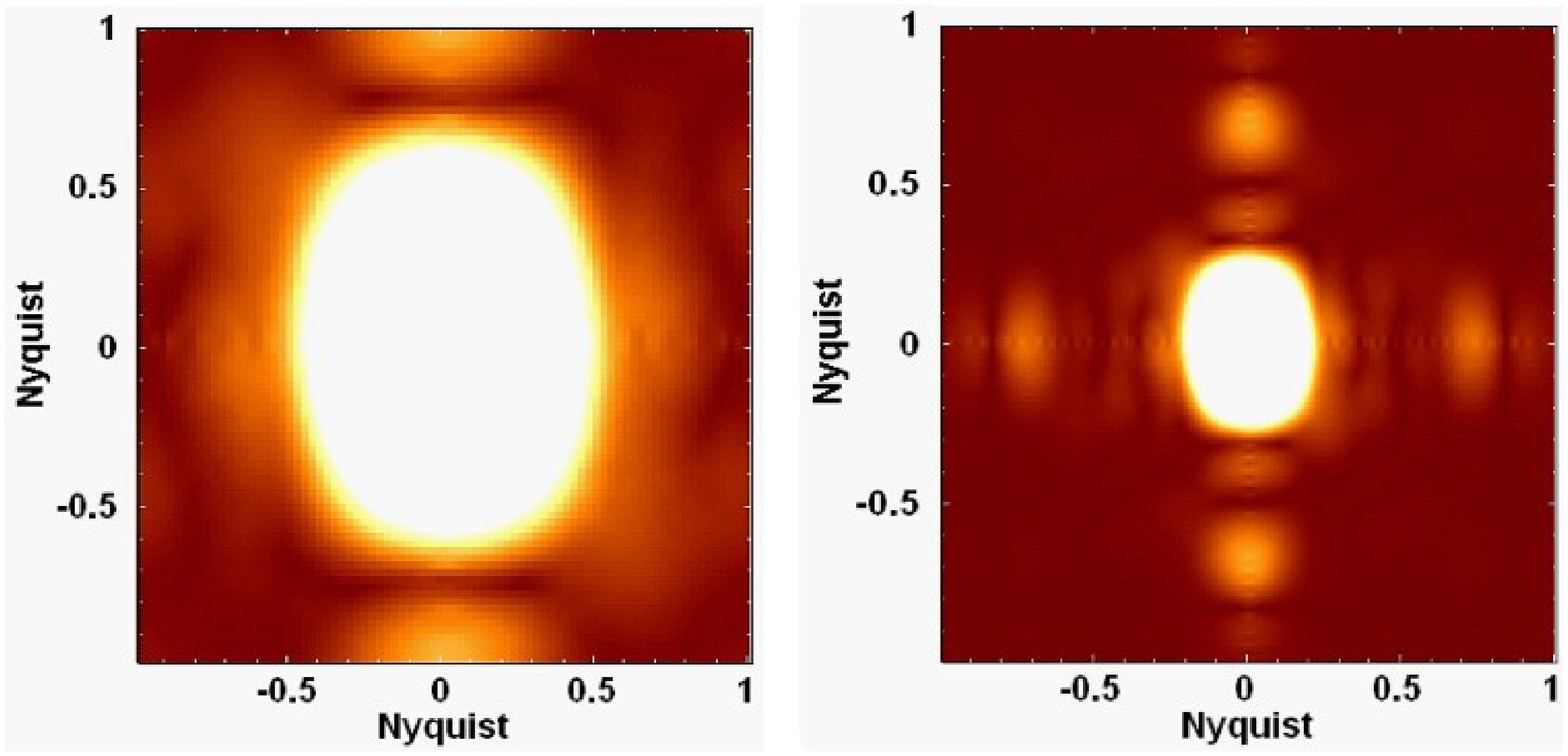}
  \caption{Left: modulus of the Fourier transform of the image of GQ Lup in Fig.~\ref{fig8}, before the spatial re-sampling. Right: modulus of the Fourier transform of the image of GQ Lup in Fig.~\ref{fig8}, after the spatial re-sampling. The high spatial-frequency components introduced by the procedure appear as bright areas farther from the centre of the image.\label{fig9}}
\end{center}
\end{figure*}

After the data reduction and the correction of the spatial displacements along the spectral axis of the data cubes, the next step in the proposed sequence for treating NIFS data cubes is the spatial re-sampling, in order to reduce the size of the spaxels. We verified that, in most cases, this procedure improves the visualization of the spatial structures. It is important to mention that the spatial re-sampling does not change the spatial resolution of the images, but only improves their appearance (by providing a clearer visualization of the contours of the structures). However, one of the main reasons for applying the spatial re-sampling to NIFS data cubes is the result obtained when this process is used together with the Richardson-Lucy deconvolution, which is the last step in the proposed sequence (see Section 7). Our previous experiences revealed that not only the final appearance of the images are better, but also the final spatial resolution is higher when the Richardson-Lucy deconvolution is applied in a spatial re-sampled data cube.

In order to apply this procedure, we use a script, written in IDL, that increases the number of spaxels in each one of the images of the data cubes, conserving the surface flux. The use of this process only, however, does not improve considerably the appearance of the images, because the re-sampled image ends up with groups of adjacent spaxels with equal or similar values. An example of that is when an image is re-sampled in order to double the number of spaxels along the horizontal and vertical axis. In that case, in order to conserve the surface flux, each spaxel of the initial image originates four others, each one of them with a value equal to one quarter of the original value. The final result is an image without significant improvements, as each group of four adjacent spaxels is visually equal to a spaxel in the initial image. In order to solve this problem, after the spatial re-sampling of an image, we also apply an interpolation in the values of the spaxels in each one of the lines and, after that, in each one of the columns of the image. It is important to emphasize that this interpolation is applied when the re-sampling is already done and only to improve the appearance of the images (removing the groups of adjacent spaxels with similar values). There are many interpolation methods that can be used here. Our previous experiences revealed that the most appropriate of these interpolation methods are: lsquadratic, quadratic and spline. The lsquadratic interpolation is performed by fitting a quadratic function, in the form $y = a + bx + cx^2$, to each group of four adjacent spaxels along a line (or a column). The quadratic interpolation is very similar to the lsquadratic interpolation, but the function $y = a + bx + cx^2$ is fitted to each group of three adjacent spaxels. The spline interpolation is performed by fitting a cubic spline to each group of four adjacent spaxels. In order to show the importance of using the interpolation after the re-sampling procedure, and also to compare the different interpolation methods, we applied the spatial re-sampling to the data cube of GQ Lup using four approaches: a simple re-sampling (without the interpolation), a re-sampling followed by a lsquadratic interpolation, a re-sampling followed by a quadratic interpolation and a re-sampling followed by a spline interpolation. Since the side effects of the non-use of the interpolation are more clearly detected when the number of spaxels in the re-sampled image is a multiple of the number of spaxels in the initial image, we applied the procedure to the data cube of GQ Lup in order to double the number of spaxels along the horizontal and vertical axis of the images. One natural question at this point is: considering that the reduced data cubes were already spatially re-sampled from the native 0.103 arcsec $\times$ 0.043 arcsec spaxels to 0.05 arcsec $\times$ 0.05 arcsec spaxels, why not to simply reduce the data cubes (with the IRAF pipeline) with smaller spaxels, instead of applying a spatial re-sampling to the data cubes ? In order to answer this question, we reduced again the data cubes of GQ Lup with spaxels of 0.025 arcsec $\times$ 0.025 arcsec and also re-applied the procedures described in Section 3. Fig.~\ref{fig7} shows graphs of the horizontal brightness profiles of the images of collapsed intermediate-wavelength intervals of the following data cubes of GQ Lup: a data cube with spaxels of 0.05 arcsec $\times$ 0.05 arcsec without the spatial re-sampling; a data cube with spaxels of 0.025 arcsec $\times$ 0.025 arcsec obtained with a simple spatial re-sampling without any interpolation; data cubes with spaxels of 0.025 arcsec $\times$ 0.025 arcsec obtained with a spatial re-sampling followed by lsquadratic, quadratic and spline interpolations; a data cube with spaxels of 0.025 arcsec $\times$ 0.025 arcsec obtained with a different data reduction procedure (which originated the spaxels of 0.025 arcsec $\times$ 0.025 arcsec) and no spatial re-sampling.

In Fig.~\ref{fig7}, we can see that a spatial re-sampling followed by an interpolation of the values provides a clearer visualization of the contours of the star in the image than a simple spatial re-sampling. Besides that, the three interpolation methods used here provided very similar results. Therefore, we conclude that any of these methods may be used in the data treatment procedure described in this paper. Finally, Fig.~\ref{fig7} also reveals that the reduction of the data cube (with the IRAF pipeline) with smaller spaxels provided a much poorer visualization of the contours of the star in the image than the re-sampling procedure followed by an interpolation.

The poorer visualization of the contours of the structures, however, is not the only disadvantage of reducing the data cubes with spaxels smaller than 0.05 arcsec $\times$ 0.05 arcsec. Another disadvantage is related with the main side effect of a spatial re-sampling: any spatial re-sampling of an image (with or without an interpolation) introduces high spatial-frequency components in the images. That is a natural consequence of this kind of procedure. If a data cube is reduced with smaller spaxels, then it will also contain a higher number of high spatial-frequency components. Such components may interfere in the process of determining the coordinates of the centres of spatial structures, which is a necessary step in the practical approach for removing the spatial displacements of structures along the spectral axis of NIFS data cubes (see Section 3). In other words, the reduction of data cubes with smaller spaxels may interfere in the correction of the DAR effect. Our previous experiences revealed that this interference only starts to be relevant if a data cube is reduced with spaxels smaller than 0.05 arcsec $\times$ 0.05 arcsec. Considering that and the poorer visualization of the structures obtained, we conclude that the most appropriate procedure for obtaining spaxels with dimensions smaller than 0.05 arcsec $\times$ 0.05 arcsec is the spatial re-sampling followed by an interpolation and not the reduction of the data cube with smaller spaxels.

One important point that is worth mentioning here is related to the signal-to-noise (S/N) ratio of the spectra of the data cubes. Although the spatial re-sampling (followed by an interpolation) improves significantly the visualization of the contours of the structures in the images, it may degrade the S/N ratio, as it is actually a second re-sampling procedure applied to the data (the first was performed during the data reduction, in order to obtain spaxels of 0.05 arcsec $\times$ 0.05 arcsec from the native 0.103 arcsec $\times$ 0.043 arcsec spaxels). In order to evaluate if this degradation of the S/N ratio is significant, we extracted a spectrum from a circular region with a radius of 0.05 arcsec, centred on the star, of the data cube of GQ Lup, before and after the spatial re-sampling. After that, we calculated the S/N ratio in the spectral range 2.2685 - 2.2800 $\mu m$ in the two spectra. The results were 113.9 and 113.5 for the spectra from the data cubes before and after the spatial re-sampling, respectively. This indicates that the spatial re-sampling reduced $\sim$ 0.4 per cent of the S/N ratio of the spectra. In our previous experiences, we have never observed a degradation higher than 1.0 per cent. Considering these very low degradations and the significant improvements obtained in the visualization of the contours of the structures, we conclude that the spatial re-sampling is a procedure that is worth applying to NIFS data cubes. 

The previous spatial re-sampling, applied to the data cube of GQ Lup in order to double the number of spaxels along the horizontal and vertical axis of the images, was only used to show the importance of the interpolation after the re-sampling (the resulting data cubes will not be used for any other purpose in this work). However, the spatial re-samplings we usually apply to NIFS data cubes result in spaxels of 0.021 arcsec $\times$ 0.021 arcsec. We choose this value for the final size of the spaxels because it is approximately a submultiple of the original dimensions of NIFS spatial pixels (0.103 arcsec $\times$ 0.043 arcsec). Using this strategy, we applied the spatial re-sampling, followed by an interpolation, to the data cube of GQ Lup (with original spaxels of 0.05 arcsec $\times$ 0.05 arcsec) and obtained a data cube with spaxels of 0.021 arcsec $\times$ 0.021 arcsec. Fig.~\ref{fig8} shows images of a collapsed intermediate-wavelength interval of the data cube of GQ Lup, before and after the spatial re-sampling.

\begin{figure*}
\begin{center}
  \includegraphics[scale=0.66]{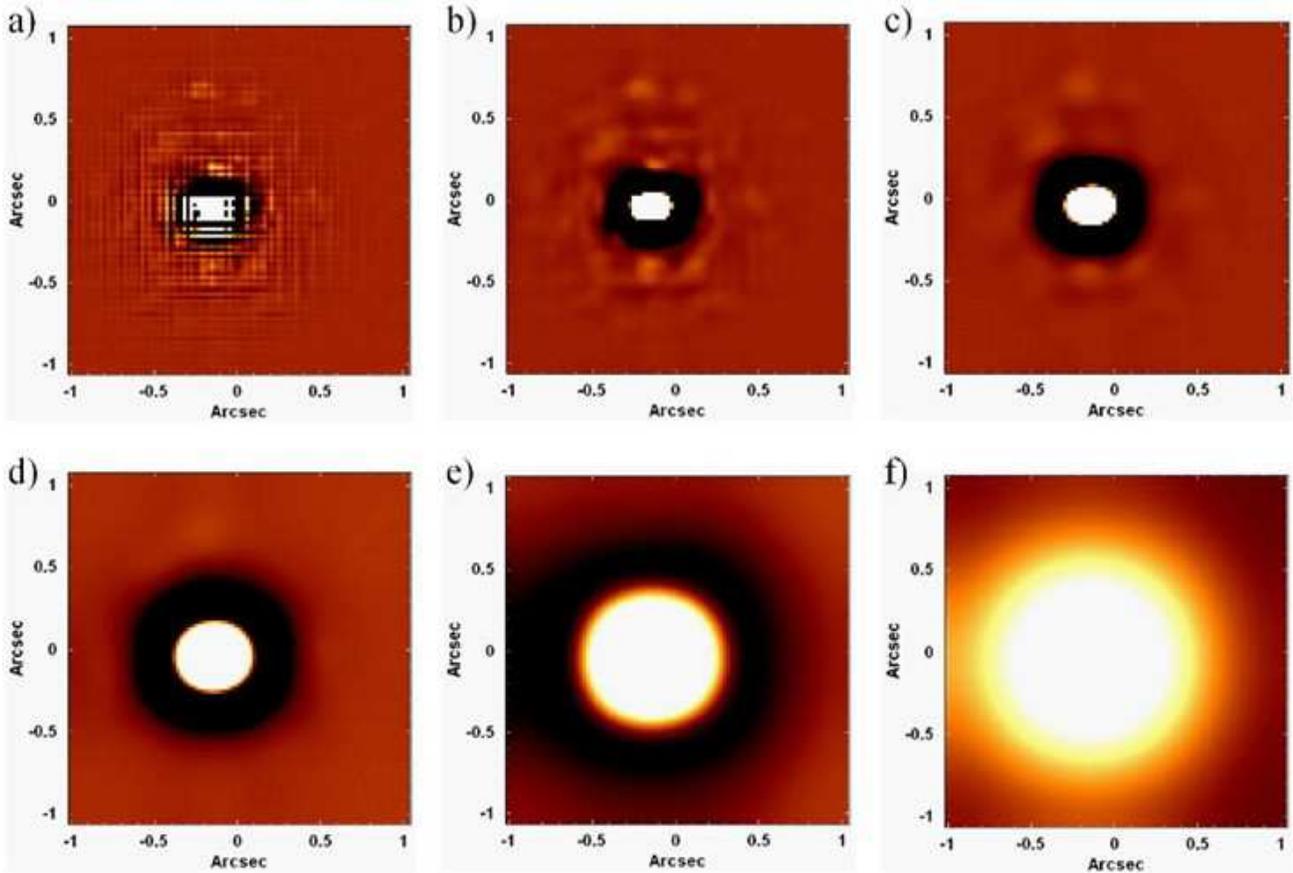}
  \caption{Images of a collapsed intermediate-wavelength interval of the data cubes a) $W_0$, b) $W_1$, c) $W_2$, d) $W_3$, e) $W_4$ and f) $W_C$, obtained with the spatial wavelet decomposition of the data cube of GQ Lup.\label{fig10}}
\end{center}
\end{figure*}

Fig.~\ref{fig8} shows that the spatial re-sampling improved the visualization of the star GQ Lup; however, as mentioned before, it also introduced high spatial-frequency components, which can be seen as thin horizontal and vertical stripes in the image (these structures are very subtle in Fig.~\ref{fig8}). It is important to mention that these high spatial-frequency components do not represent a considerable problem here, as they are almost entirely removed by the Butterworth spatial filtering (see Section 5). 

In order to determine the high-frequency components introduced by the spatial re-sampling, we calculated the Fourier transforms of the two images in Fig.~\ref{fig8}. Fig.~\ref{fig9} shows the moduli of the Fourier transforms of the images in Fig.~\ref{fig8}.

In Fig.~\ref{fig9}, the high-frequency components introduced by the spatial re-sampling appear very clearly as bright areas farther from the centre of the image. Most of these components have spatial frequencies higher than 0.5 Ny (where Ny represents the Nyquist frequency, which corresponds to half of the sampling-frequency of the image). It is important to mention that the definition of the Nyquist frequency in the images in Fig.~\ref{fig9}, and elsewhere in this paper, is related to the spaxel sampling of each image. Therefore, according to this definition, a complete oscillation within two spaxels has a frequency of 1 Ny. 

\begin{figure*}
\begin{center}
  \includegraphics[scale=0.66]{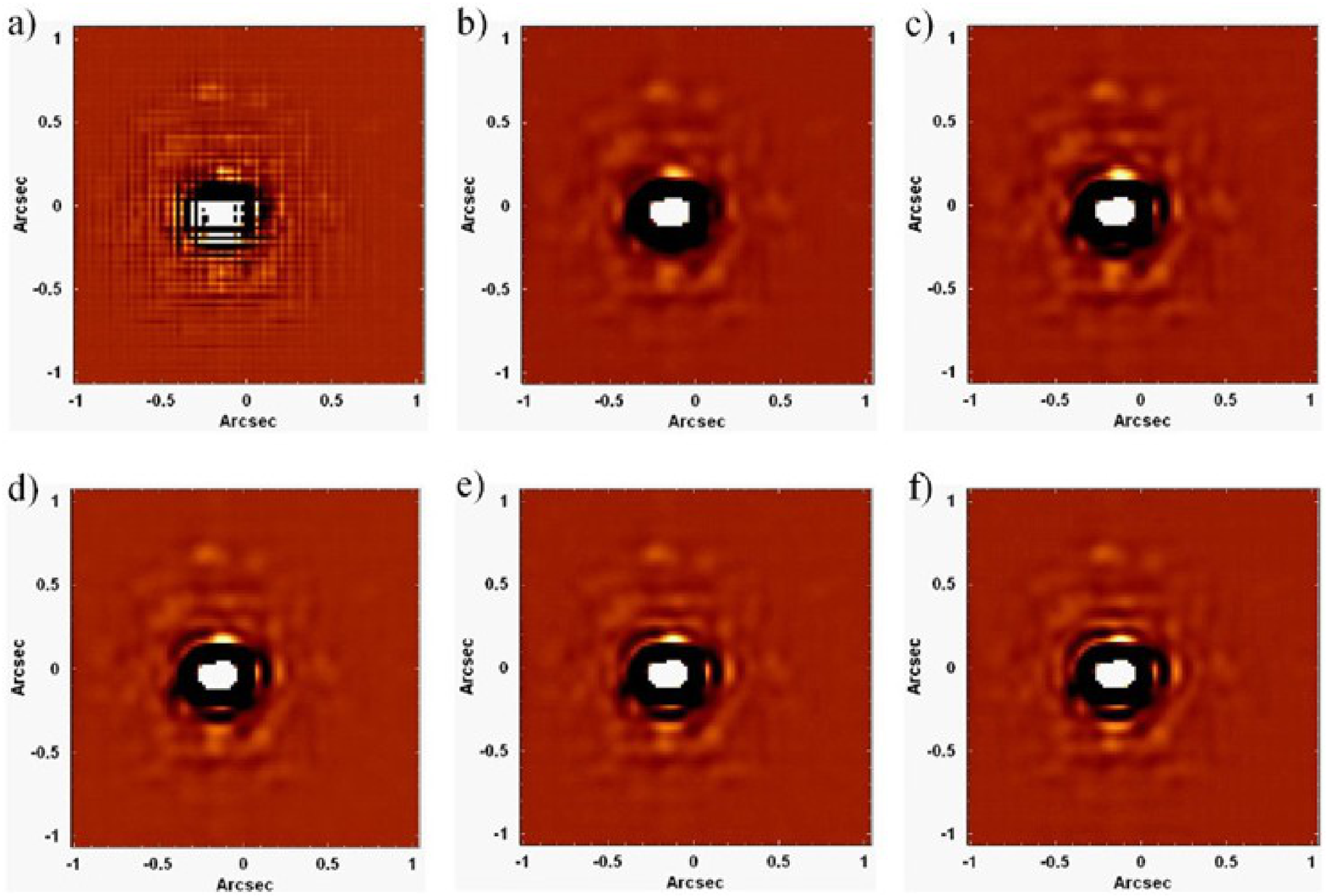}
  \caption{Images of a collapsed intermediate-wavelength interval of the $W_0$ data cube of GQ Lup, a) before the Butterworth spatial filtering and after the Butterworth spatial filtering with b) $n = 2$, c) $n = 3$, d) $n = 4$, e) $n = 5$ and f) $n = 6$.\label{fig11}}
\end{center}
\end{figure*}

\section{Butterworth spatial filtering}

The next step in the proposed sequence for treating NIFS data cubes is the Butterworth spatial filtering \citep{gon02}, which consists of a filtering process performed directly in the frequency domain. It is well known that a Fourier transform of a function passes this function to the frequency domain and allows one to analyse its frequency components. As a consequence, using the Fourier transform, it is possible to modify the frequency components of a function. A filtering process performed in the frequency domain is based on the idea of calculating the Fourier transform of a function and eliminating certain frequency components. After the removal of the frequency components is done, an inverse Fourier transform of the function is calculated, completing the process.

In the case of the Butterworth spatial filtering of images, the main steps of the procedure can be summarized as: calculation of the Fourier transform ($F(u,v)$) of the image; multiplication of the Fourier transform $F(u,v)$ by the image corresponding to the Butterworth filter ($H(u,v)$); calculation of the inverse Fourier transform of the product $F(u,v) \cdot H(u,v)$; extraction of the real part of the calculated inverse Fourier transform. Since the original image and the filter $H(u,v)$ have only real values, the inverse Fourier transform of $F(u,v) \cdot H(u,v)$ is expected to be real as well; however, due to computational uncertainties, such an inverse Fourier transform often shows a low modulus imaginary component. That is the reason why the last step of the procedure is the extraction of the real part of the inverse Fourier transform of $F(u,v) \cdot H(u,v)$. Since the purpose of the filtering process in this work is the removal of high spatial-frequency noise, only low-pass filters were used.

\begin{figure*}
\begin{center}
  \includegraphics[scale=0.66]{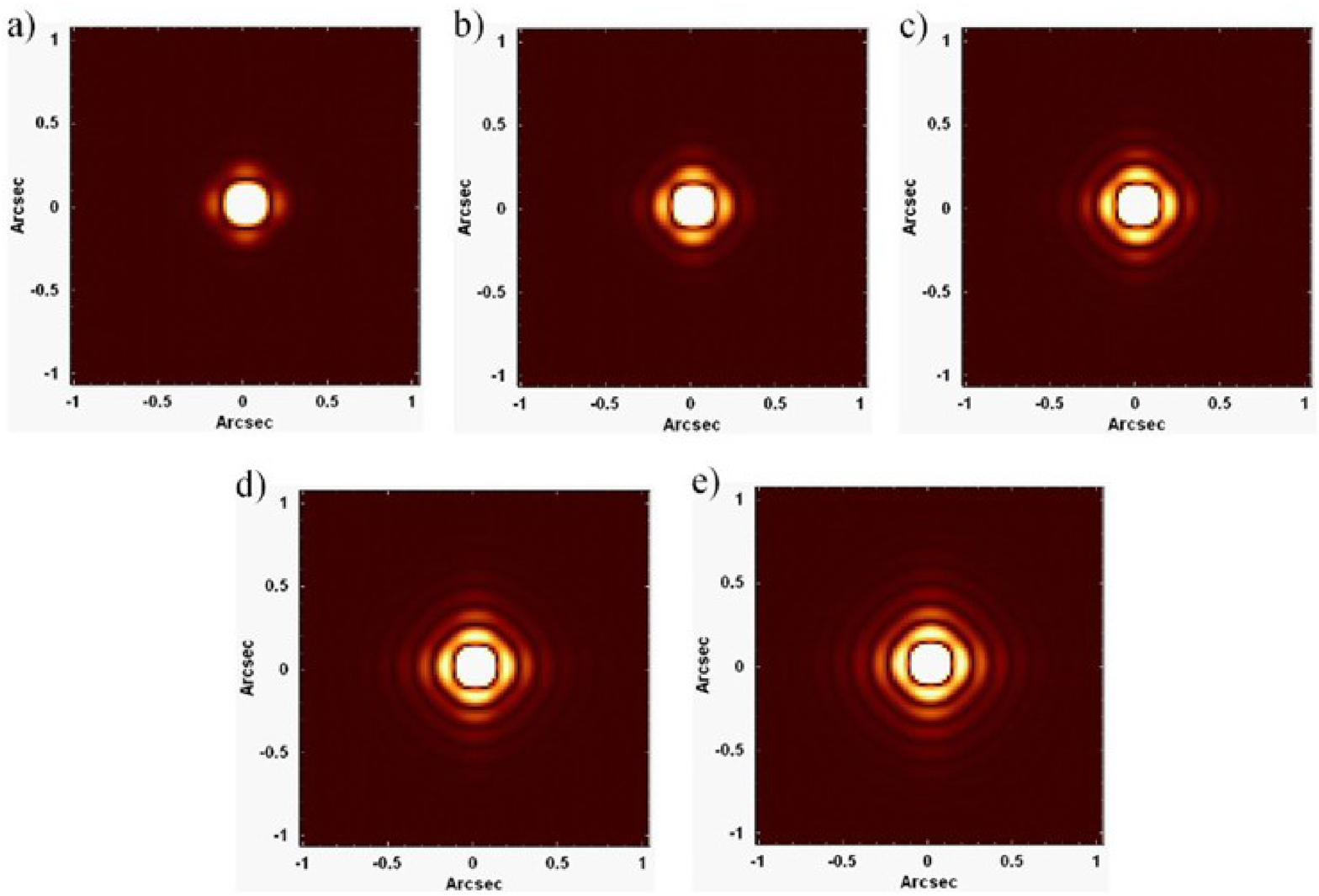}
  \caption{Moduli of the inverse Fourier transforms of Butterworth filters with cut-off frequency of 0.39 Ny and order a) $n = 2$, b) $n = 3$, c) $n = 4$, d) $n = 5$ and e) $n = 6$.\label{fig12}}
\end{center}
\end{figure*}

In a low-pass Butterworth spatial filtering of order $n$, the filter $H(u,v)$ is given by

\begin{equation}
H\left(u,v\right) = \frac{1}{1+\left(\frac{D\left(u,v\right)}{D_0}\right)^{2n}},
\end{equation}
where $D(u,v)$ is the distance between the points $(u,v)$ and $(u_0,v_0)$ in the frequency domain and $D_0$ (the cut-off frequency) is the distance from $(u_0,v_0)$ at which $H(u,v) = 0.5$. In all these definitions, the point $(u_0,v_0)$ is located at the centre of the image corresponding to the Fourier transform. In equation (7), we can see that the higher is the value of $n$, the steeper is the ``cut'' made by the filter in the frequency domain.

In order to apply the Butterworth spatial filtering to data cubes, we use a script, written in IDL, that performs all the steps of the filtering in each one of the images of the data cubes. In the case of NIFS data cubes, considering the typical morphology of the Fourier transforms of the images (analogous to the one shown in Fig.~\ref{fig9}), we verified that the most adequate filter to be used in the filtering process is given by

\begin{equation}
\begin{split}
H\left(u,v\right) = \left\{\frac{1}{1 + \left[\sqrt{\left(\frac{u - u_0}{a}\right)^2 + \left(\frac{v - v_0}{b}\right)^2}\right]^{2n}}\right\} \times \\ 
\times \left\{\frac{1}{1 + \left[\frac{\vert u - u_0 \vert}{c}\right]^{2n}} \times \frac{1}{1 + \left[\frac{\vert v - v_0 \vert}{d}\right]^{2n}}\right\},
\end{split}
\end{equation}
which corresponds to the product of a filter with an elliptical shape by another with a rectangular shape. In equation (8), $a$ and $b$ are the cut-off frequencies along the horizontal and vertical axis, respectively, of the elliptical filter, and $c$ and $d$ are the cut-off frequencies along the horizontal and vertical axis, respectively, of the rectangular filter. It is easy to see that the shape of the filter given by equation (8) is very similar to the shape of the region corresponding to the low frequencies in the Fourier transforms of the images of NIFS data cubes (Fig.~\ref{fig9}). Our previous experiences showed that, in most cases, good results are obtained if we simply assume $a = b = c = d$. This strategy makes it easier and faster to determine the most appropriate values for the cut-off frequencies of the observations.

An analysis of the Fourier transform in Fig.~\ref{fig9} revealed that an appropriate value for the cut-off frequency to be used in the Butterworth spatial filtering of the data cube of GQ Lup is approximately 0.39 Ny. However, in order to determine the most adequate value of $n$ to be used, first of all, we applied a spatial wavelet decomposition \citep{sta06} to the data cube. A wavelet decomposition is a process in which a signal is decomposed in many components with different characteristic frequencies. In this case, we used this procedure to decompose each image of the data cube of GQ Lup in five wavelet components ($w_0$, $w_1$, $w_2$, $w_3$ and $w_4$) and $w_c$, which corresponds to a smooth version of the original image. In this work, to perform this spatial wavelet decomposition, we used the \`A Trous algorithm, which has two important characteristics:

\begin{itemize}

\item the sum of the components $w_0$, $w_1$, $w_2$, $w_3$, $w_4$ and $w_c$ obtained for each image, corresponds to the original image,

\item the average of each wavelet component is 0.

\end{itemize}

Since this wavelet decomposition was applied to each image of the data cube, at the end of the process, we obtained six data cubes ($W_0$, $W_1$, $W_2$, $W_3$, $W_4$ and $W_C$). In this notation, $w_0$ and $w_4$ represent the wavelet components with the highest and lowest spatial frequencies, respectively. Fig.~\ref{fig10} shows images of a collapsed intermediate-wavelength interval of the data cubes $W_0$, $W_1$, $W_2$, $W_3$, $W_4$ and $W_C$, obtained with the wavelet decomposition of the data cube of GQ Lup. We can see that $W_0$ reveals the existence of a considerable amount of high spatial-frequency noise, which appear in the form of thin horizontal and vertical stripes across the image. These structures are so abundant that we can barely see the star in the image of $W_0$. Some of these stripes may be associated with instrumental noise; however, it is probable that part of them correspond to the high-frequency components introduced by the spatial re-sampling procedure (Section 4).

After that, we applied the Butterworth spatial filtering procedure to $W_0$ (with a cut-off frequency of 0.39 Ny) with $n = 2$, $n = 3$, $n = 4$, $n = 5$ and $n = 6$. Fig.~\ref{fig11} shows images of a collapsed intermediate-wavelength interval of the $W_0$ data cube, before and after the filtering process with these values of $n$.

Fig.~\ref{fig11} reveals that the Butterworth spatial filtering improved considerably the visualization of the star GQ Lup, which was barely visible in the original image of $W_0$. However, the data cubes filtered with orders higher than 2 show structures similar to concentric rings. These structures were introduced by the filtering process and start to be visible with $n = 3$ (although possible traces of an inner ring can be seen in the image corresponding to $n = 2$), but become more and more intense as $n$ grows. This is an expected consequence of the use of the Butterworth spatial filtering with high orders. In order to explain this behaviour, first of all, it is important to mention that, according to the convolution theorem, the product between two functions in the frequency domain is equivalent to the convolution between these functions in the spatial domain. In other words, the convolution between two functions is equivalent to the product between the Fourier transforms of these functions. Therefore, in the case of the Butterworth spatial filtering of a data cube, the product between the Fourier transform of each image of the data cube and the image corresponding to the Butterworth filter is equivalent to the convolution between each original image of the data cube and the inverse Fourier transform of the Butterworth filter. Fig.~\ref{fig12} shows the moduli of the inverse Fourier transforms of the Butterworth filters used to generate each image in Fig.~\ref{fig11}.

In Fig.~\ref{fig12}, the inverse Fourier transform of the filter with $n = 6$ shows a series of concentric rings that decrease in number and intensity as $n$ decreases. The inverse Fourier transform of the filter with $n = 2$ shows only one weak ring (which is almost undetectable). Therefore, considering the behaviour of the images in Fig.~\ref{fig12} and the fact that the Butterworth spatial filtering of a data cube is equivalent to the convolution between the inverse Fourier transform of the Butterworth filter and the original images of the data cube, we can see why the $W_0$ data cubes of GQ Lup filtered with high values of $n$ show structures similar to concentric rings. In Fig.~\ref{fig11}, we can also see a considerable number of relatively small structures around GQ Lup. Most of these structures are artefacts introduced by the AO in the observation. The only exception is the farthest source at west of GQ Lup, which corresponds to GQ Lup b, a brown dwarf already analysed in previous studies \citep{lav09}.

One important point to be mentioned is that the Fourier transform of a function has a periodicity, due to the form by which it is defined, and a convolution between functions with some periodicity generates results different from a convolution between functions that do not have this property. This fact can cause problems in a Butterworth spatial filtering, as this procedure, according to what was mentioned before, is equivalent to a convolution between the original images and the inverse Fourier transform of the Butterworth filter. In order to solve this issue, we apply a `padding' procedure. We will not describe this process here, but more details can be found in \citet{gon02}.

The rings observed in Fig.~\ref{fig11} could be detected so clearly only because we analysed images of the $W_0$ data cube of GQ Lup. In images of original NIFS data cubes (without any wavelet decomposition), it is much harder to see such structures, due to the fact that they are immersed in the low-frequency components of the PSFs of the objects, being obfuscated by these. It is worth mentioning that, due to their nature, these concentric rings are only visible around point-like sources.

\begin{figure*}
\begin{center}
  \includegraphics[scale=0.52]{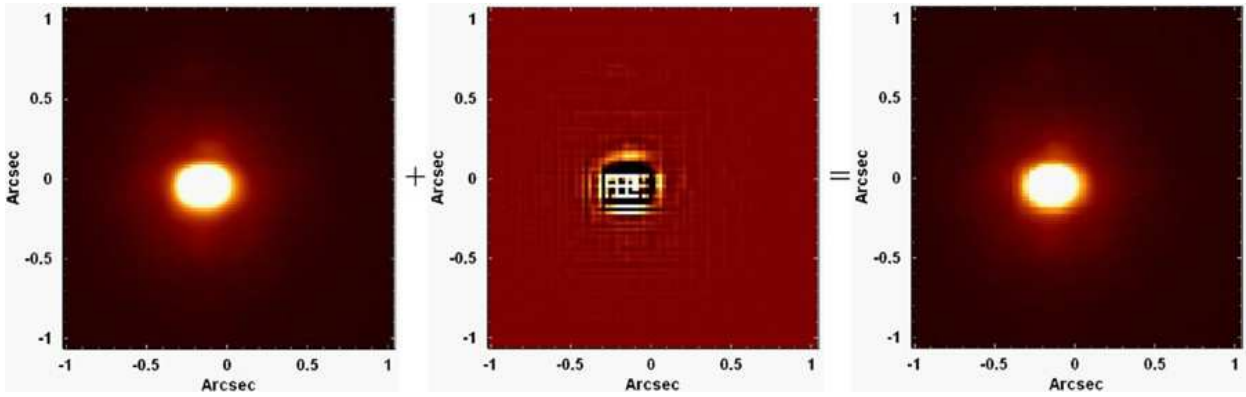}
  \caption{Left: image of a collapsed intermediate-wavelength interval of the data cube of GQ Lup, after the Butterworth spatial filtering. Right: image of the same collapsed intermediate-wavelength interval of the data cube of GQ Lup, before the Butterworth spatial filtering. Centre: image of a collapsed intermediate-wavelength interval of the data cube corresponding to the difference between the previous two.\label{fig13}}
\end{center}
\end{figure*}

\begin{figure*}
\begin{center}
  \includegraphics[scale=0.52]{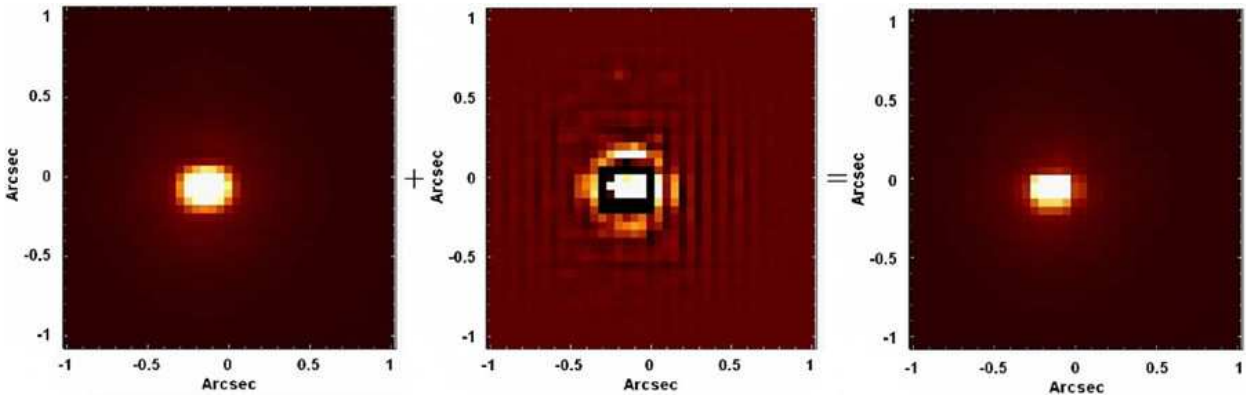}
  \caption{Left: image of a collapsed intermediate-wavelength interval of the data cube of GQ Lup with spaxels of 0.05 arcsec, after the Butterworth spatial filtering. Right: image of the same collapsed intermediate-wavelength interval of the data cube of GQ Lup with spaxels of 0.05 arcsec, before the Butterworth spatial filtering. Centre: image of a collapsed intermediate-wavelength interval of the data cube corresponding to the difference between the previous two.\label{fig14}}
\end{center}
\end{figure*}

\begin{figure*}
\begin{center}
  \includegraphics[scale=0.52]{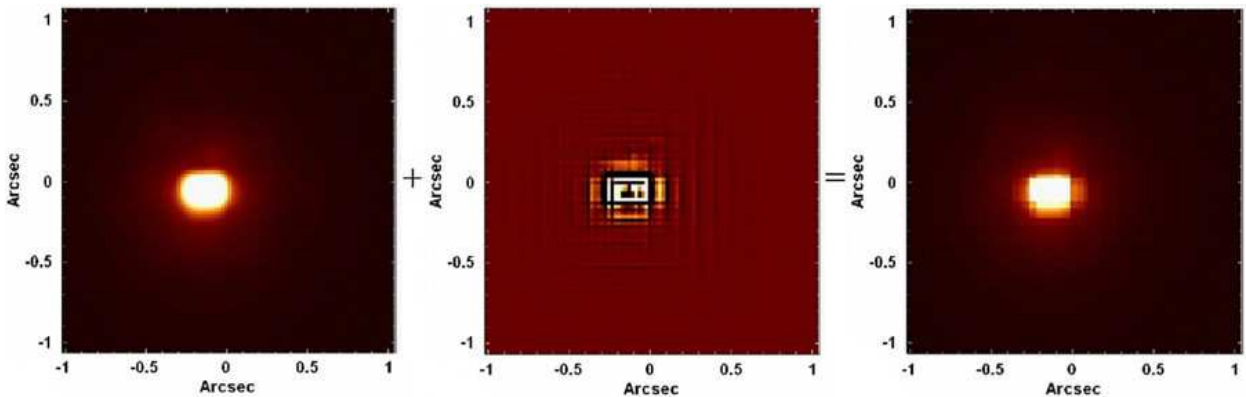}
  \caption{Left: image of a collapsed intermediate-wavelength interval of the data cube of GQ Lup spatially re-sampled (to spaxels of 0.021 arcsec) without any interpolation, after the Butterworth spatial filtering. Right: image of the same collapsed intermediate-wavelength interval of the data cube of GQ Lup spatially re-sampled (to spaxels of 0.021 arcsec), before the Butterworth spatial filtering. Centre: image of a collapsed intermediate-wavelength interval of the data cube corresponding to the difference between the previous two.\label{fig15}}
\end{center}
\end{figure*}

Fig.~\ref{fig11} shows that the Butterworth spatial filtering with $n = 2$ removed most of the high spatial-frequency noise without introducing significant rings in the images. The explanation given above states that this is a mathematical consequence of the Butterworth filtering and not only a particular result obtained with the data cube of GQ Lup. Therefore, we conclude that a Butterworth spatial filtering with $n=2$ is the most appropriate to be applied to NIFS data cubes.

The cut-off frequency of the filtering process (0.39 Ny in the case of GQ Lup) of NIFS data cubes changes considerably depending on the observed object. We have observed (in previous experiences) values ranging from 0.19 Ny to more than 0.40 Ny. This significant variation in the cut-off frequency is caused by different seeing conditions of the observations, together with the fact that the AO correction of this instrument does not have the same efficiency for all objects. A good strategy to determine the most appropriate cut-off frequency for the Butterworth spatial filtering of a data cube is to perform a wavelet decomposition of the original data cube and then apply the filtering procedure, with different cut-off frequencies, to the $W_0$ data cube. Since the $w_0$ wavelet component has the highest spatial-frequency components, performing the filtering procedure on $W_0$ will reveal very clearly the amount of noise removed and possible distortions or artefacts introduced by the process. An appropriate Butterworth spatial filtering must remove the greatest possible amount of noise, without affecting the PSF of the object.

We applied the Butterworth spatial filtering to the original data cube (without any wavelet decomposition) of GQ Lup, using a filter given by equation (8), a cut-off frequency of 0.39 Ny, and $n = 2$. Fig.~\ref{fig13} shows images of a collapsed intermediate-wavelength interval of the following data cubes: the original data cube of GQ Lup, the filtered data cube of GQ Lup, and the data cube corresponding to the difference between the previous two.

\begin{figure*}
\begin{center}
  \includegraphics[scale=0.40]{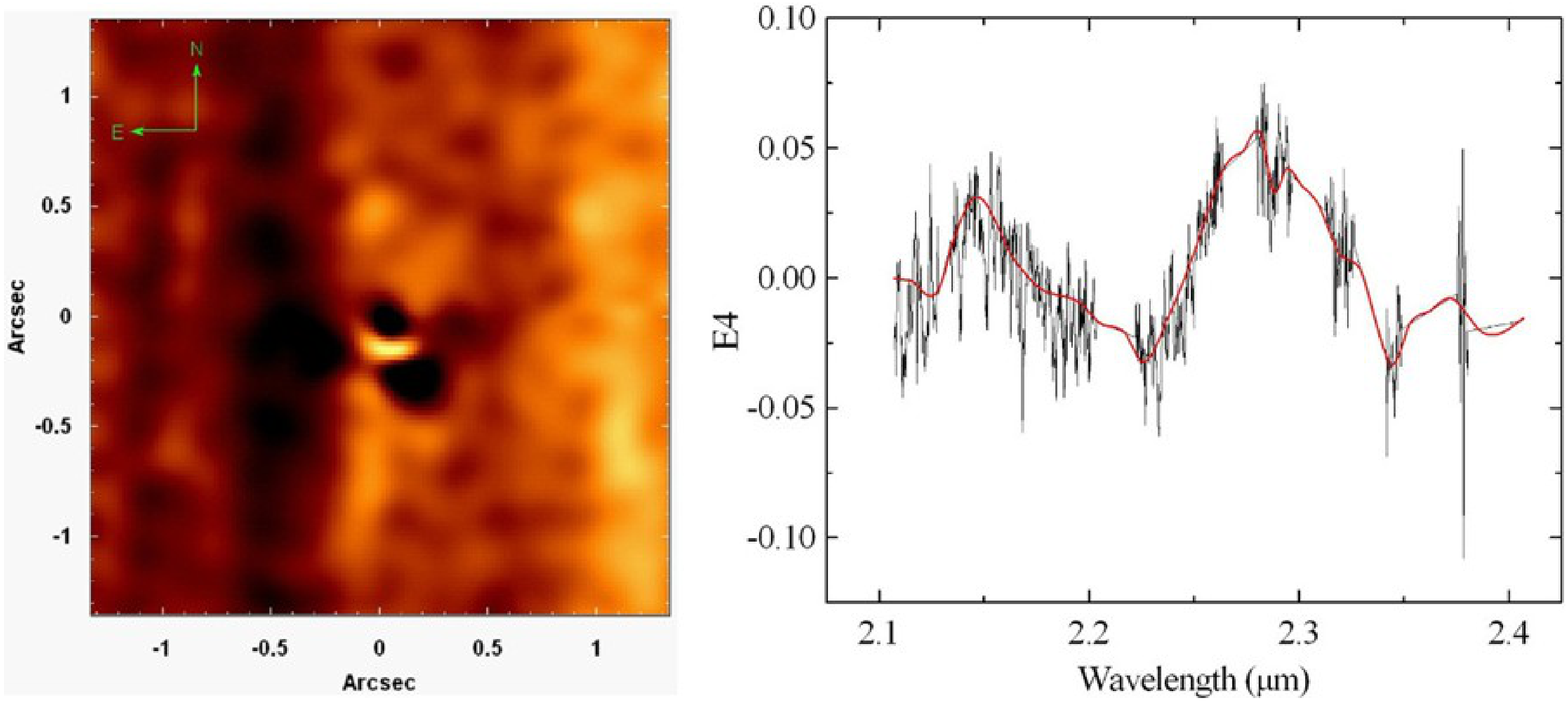}
  \caption{Tomogram and eigenspectrum (together with the fitted spline, shown in red) corresponding to eigenvector \textit{E}4, obtained with PCA Tomography of the data cube of M87, after the removal of the spectral lines.\label{fig16}}
\end{center}
\end{figure*}

Fig.~\ref{fig13} reveals that the Butterworth spatial filtering removed a considerable amount of high-frequency noise (in the form of thin vertical and horizontal stripes) from the images of the original data cube of GQ Lup. One important question at this point is: were the high spatial-frequency components in Fig.~\ref{fig13} entirely introduced by the spatial re-sampling procedure or were at least part of them already present in the data cube before the spatial re-sampling ? In order to answer this question, we applied the Butterworth spatial filtering to the data cube of GQ Lup before the spatial re-sampling. Fig.~\ref{fig14} shows images analogous to the ones in Fig.~\ref{fig13} but from the data cube of GQ Lup before the spatial re-sampling. In order to evaluate also the influence of the interpolation (applied after the spatial re-sampling) on the high spatial-frequency components in Fig.~\ref{fig13}, we applied the Butterworth spatial filtering to the data cube of GQ Lup after the spatial re-sampling (to spaxels of 0.021 arcsec $\times$ 0.021 arcsec) without any interpolation. Fig.~\ref{fig15} shows images analogous to the ones in Fig.~\ref{fig13} and in Fig.~\ref{fig14} but from the data cube of GQ Lup after the spatial re-sampling without any interpolation.

In Fig.~\ref{fig14}, we can see high spatial-frequency components (in the form of horizontal and vertical stripes across the image) similar to the ones observed in Fig.~\ref{fig13}, although the spatial sampling is different. This indicates that at least part of the high spatial-frequency components observed in Fig.~\ref{fig13} were already present in the data cube with spaxels of 0.05 arcsec $\times$ 0.05 arcsec and were not introduced by the spatial re-sampling procedure. Fig.~\ref{fig15} reveals high spatial-frequency components almost identical to the ones observed in Fig.~\ref{fig13}. This indicates that, although the spatial re-sampling has introduced high spatial-frequency components in the images, the interpolation applied had little effect on them.

\section{`Instrumental fingerprint' removal}

We have observed that many of the NIFS data cubes we have previously analysed showed certain `instrumental fingerprints' that, usually, presented the form of large vertical stripes in the images. Such structures also had very specific spectral signatures. The removal of instrumental fingerprints from NIFS data cubes involves the use of the PCA Tomography technique \citep{ste09}.

\begin{figure*}
\begin{center}
  \includegraphics[scale=0.37]{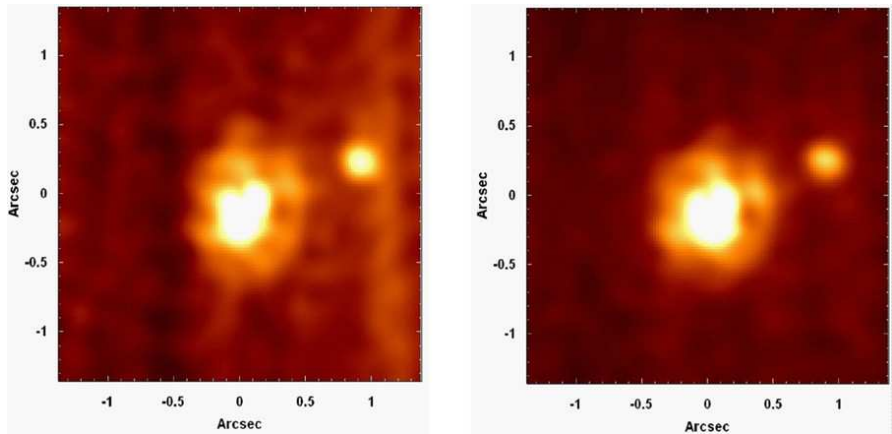}
  \caption{Left: subtraction of the image of the collapsed wavelength interval 2.1820 - 2.1911 $\mu m$ from the image of the collapsed wavelength interval 2.2510 - 2.2600 $\mu m$, both from the data cube of M87 before the instrumental fingerprint removal. Right: the same image shown at left, but from the data cube of M87 after the instrumental fingerprint removal.\label{fig17}}
\end{center}
\end{figure*}

PCA is a statistical technique often used to analyse multidimensional data sets. This method is efficient to extract information from such a large amount of data because it allows one to identify patterns and correlations in the data that, otherwise, would probably not be detected. Another use of PCA is its ability to reduce the dimensionality of the data, without significant loss of information. This is very useful to handle data sets with considerably large dimensions. Mathematically, PCA is defined as an orthogonal linear transformation of coordinates that transforms the data to a new system of (uncorrelated) coordinates, such that the first of these new coordinates (eigenvector \textit{E}1) explains the highest fraction of the data variance, the second of these new coordinates (eigenvector \textit{E}2) explains the second highest fraction of the data variance and so on. The coordinates generated by PCA are called principal components and it is important to mention that, by construction, all of them are orthogonal to each other (and therefore uncorrelated - for more details, see Murtagh \& Heck 1987; Fukunaga 1990).

PCA Tomography is a method that consists of applying PCA to data cubes. In this case, the variables correspond to the spectral pixels of the data cube and the observables correspond to the spaxels of the data cube. Since the eigenvectors are obtained as a function of the wavelength, they have a shape similar to spectra and, therefore, we call them eigenspectra. On the other hand, since the observables are spaxels, their projections on the eigenvectors are images, which we call tomograms.

In order to remove instrumental fingerprints from an NIFS data cube, first of all, we remove its spectral lines. This step is necessary because the spectral signature of the instrumental fingerprint is associated only with the continuum and not with the spectral lines; therefore, the removal of these lines makes the detection of the problem easier. We then apply the PCA Tomography technique to the data cube without the spectral lines and select the obtained eigenvectors that are related to the fingerprint. It is important to mention that, if PCA Tomography is applied to the data cube with spectral lines, it is probable that the obtained eigenvectors that are related to the fingerprint have also some relation to other phenomena associated with spectral lines. In other words, it would be more difficult to isolate the fingerprint in specific eigenvectors without the `contamination' of them by spectral lines. After the selection of the eigenvectors related to the fingerprint, we fit splines to the corresponding eigenspectra and construct a data cube using the tomograms associated with these eigenspectra and the fitted splines (more details about the reconstruction of data cubes using tomograms and eigenspectra can be found in Steiner et al. 2009). The result of this process is a data cube containing, only, the instrumental fingerprint. Finally, we subtract the data cube containing the fingerprint from the original one. All previous steps are performed using scripts written in IDL. More details about the detection and removal of instrumental fingerprints from data cubes will be discussed in Steiner et al. (in preparation). We applied the procedure for instrumental fingerprint removal to the data cube of the nuclear region of the galaxy M87, which is described in Section 2. This data cube was treated with all the techniques discussed in the previous sections. Fig.~\ref{fig16} shows the tomogram and the eigenspectrum (together with the fitted spline) corresponding to eigenvector \textit{E}4, obtained with PCA Tomography of the data cube of M87, after the removal of the spectral lines. Eigenvector \textit{E}4 explains  $\sim 6.0 \times 10^{-4}$ per cent of the variance of the data cube without the emission lines. 

In Fig.~\ref{fig16}, we can see that the spectral signature of the instrumental fingerprint in the data cube of M87 has a low-frequency pattern, which is essentially the same in all NIFS data cubes, in the \textit{K} band, we have analysed so far. In data cubes obtained in the \textit{Z}, \textit{J} and \textit{H} bands, the spatial morphology and the spectral signature of the fingerprint are also very similar to what is shown in Fig.~\ref{fig16}. We used the data cube of M87 to show the fingerprint typically found in this instrument because this artefact is better visualized in data cubes of extended objects. In data cubes of point-like sources (like GQ Lup, for example), the fingerprint is detected in a much more diffuse way and, in some cases, may not be detected at all. In cases like that, the procedure for fingerprint removal is, obviously, not necessary. 

\begin{figure*}
\begin{center}
  \includegraphics[scale=0.25]{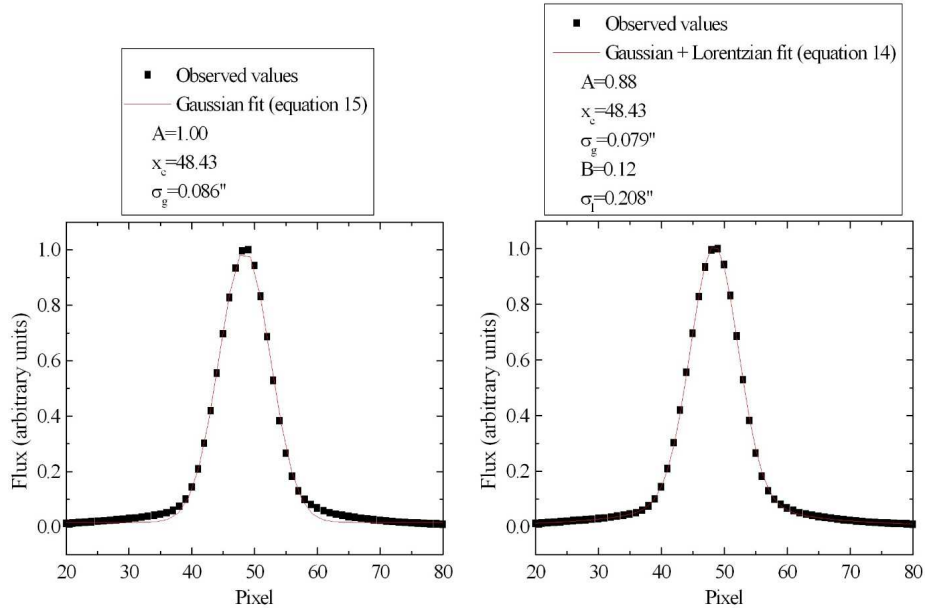}
  \caption{Left: horizontal brightness profile of the image of a collapsed intermediate-wavelength interval of the data cube of GQ Lup, with a simple Gaussian fit. Right: same brightness profile, with a Gaussian + Lorentzian fit.\label{fig18}}
\end{center}
\end{figure*}

Since eigenvector \textit{E}4 in Fig.~\ref{fig16} explains only $\sim 6.0 \times 10^{-4}$ per cent of the variance of the data cube without the emission lines, one could think that this effect is too weak to affect significantly any scientific analysis to be performed. However, that is not the case. In order to show that, we elaborated images of two collapsed wavelength intervals of the data cube of M87 before the instrumental fingerprint removal: 2.2510 - 2.2600 $\mu m$ and 2.1820 - 2.1911 $\mu m$. Then, we subtracted the second image from the first one. This same procedure was also applied to the data cube of M87 after the instrumental fingerprint removal. The results are shown in Fig.~\ref{fig17}. The instrumental fingerprint clearly affects the image without the correction. However, we can also see that our procedure was so effective that left almost no traces of the fingerprint. The reason why this fingerprint appeared so clearly in an image resulting from the subtraction of two continuum images is that, as explained above (and as Fig.~\ref{fig16} shows), this effect is associated with the continuum and not with the spectral lines. A subtraction of continuum images in different spectral regions in a data cube may be performed to evaluate differences in the inclination of the continuum in different spatial regions. In the case of Fig.~\ref{fig17}, for example, the bright regions have redder spectra than the dark ones, which may be due to the presence of thermal emission from dust in the formers.  There are, however, other studies of the spectral continuum that may be affected by the instrumental fingerprint. One example is the spectral synthesis. Considering all of that, we conclude that the instrumental fingerprint in NIFS data cubes may affect significantly the analysis in some studies and, therefore, should be removed in these cases.

The origin of the instrumental fingerprint in NIFS data cubes is still uncertain. NIFS reads out through four amplifiers in 512 pixel blocks in the spectral direction. Therefore, slight bias level differences between the four amplifiers could be one possible explanation for this fingerprint. If we try to evaluate any periodicity in the spectral behaviour revealed by Fig.~\ref{fig16} (despite the irregularities of this behaviour), we obtain a value of $\sim$ 0.1 $\mu m$. Such a periodicity in a full spectrum of $0.4$ $\mu m$ supports the idea of bias level differences between the four amplifiers generating the fingerprint. However, we cannot exclude other possibilities, like the fingerprint being generated by some step during the data reduction. Although we do not know the origin of the instrumental fingerprint, the method we present here has proved to be very effective in removing this effect from NIFS data cubes.

\section{Richardson-Lucy deconvolution}

The final step in the proposed sequence for treating NIFS data cubes is the Richardson-Lucy deconvolution \citep{ric72, luc74}. It is well known that the image of an astronomical object observed from the Earth's surface corresponds to the convolution between the original image of the object and the PSF due to the Earth's atmosphere and to the instrument. Therefore, considering also the existence of instrumental noise in the observation, the observed image of an object ($I(x,y)$) can be given by

\begin{equation}
I\left(x,y\right) = \left(P*O\right)\left(x,y\right) + N\left(x,y\right),
\end{equation}
where $O(x,y)$ is the original image of the object, $P(x,y)$ is the PSF of the observation, and $N(x,y)$ is an instrumental additive noise.

A deconvolution is an iterative process whose purpose is to revert the effect of a convolution. In the astronomical case, the deconvolution problem is usually put in the following way: knowing the observed image $I(x,y)$ and the PSF of the observation $P(x,y)$, one wishes to obtain the original image $O(x,y)$.

There are different deconvolution procedures that can be used in astronomical images, like Maximum Entropy, Van Cittert, Landweber, and Richardson-Lucy (all of these procedures are discussed in Starck \& Murtagh 2006). However, our tests showed that the Richardson-Lucy provides the best results in NIFS data cubes, as the other methods compromise the visualization of faint structures or even introduce a number of artefacts to the images. Therefore, we decided to use the Richardson-Lucy deconvolution in this work. This process is given by the following iterative equation:

\begin{equation}
O^{n+1}\left(x,y\right) = \left[\frac{I\left(x,y\right)}{\left(P*O\right)\left(x,y\right)} * P^t\left(x,y\right)\right] \times O^n\left(x,y\right),
\end{equation}
where $P^t(x,y)$ is the transpose of $P(x,y)$. The procedure is applied to each image of a data cube using a script written in IDL.

The use of AO (see Davies \& Kasper 2012 for a recent review) has the purpose of making the spatial resolution of observations as close as possible to the diffraction limit, which is given by the Airy function:

\begin{figure*}
\begin{center}
  \includegraphics[scale=0.42]{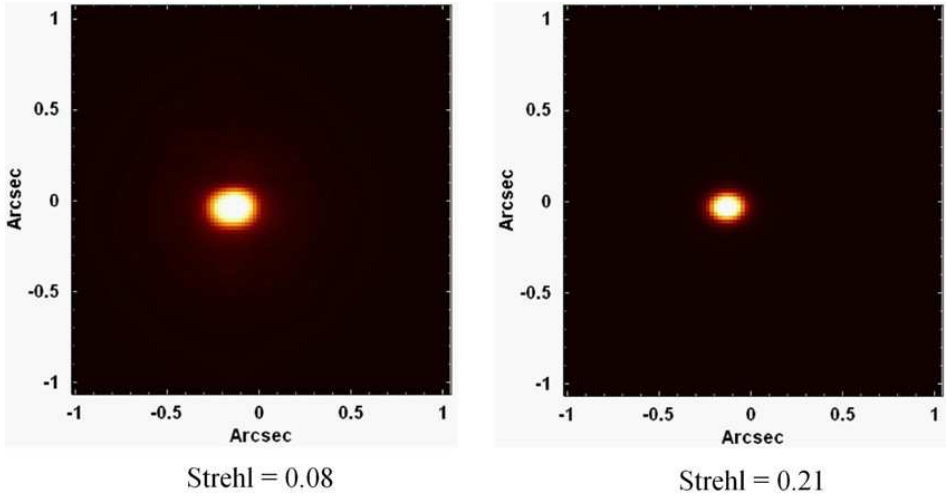}
  \caption{Left: image of a collapsed intermediate-wavelength interval of the data cube of GQ Lup, before the Richardson-Lucy deconvolution. Right: image of the same collapsed intermediate wavelength-interval of the data cube of GQ Lup, after the Richardson-Lucy deconvolution.\label{fig19}}
\end{center}
\end{figure*}

\begin{equation}
I\left(\theta\right) = I\left(0\right) \times 4 \times \left[\frac{J_1\left(k \times a \times sin \theta\right)}{k \times a \times sin \theta}\right]^2,
\end{equation}
where $J_1$ is the Bessel function of order 1, $a$ is the radius of the circular opening that caused the diffraction, and $k$ is the wavenumber, which is given by

\begin{equation}
k = \frac{2\pi}{\lambda}.
\end{equation}

However, this technique is not perfect and, at the end of the process, one generally obtains a PSF with two components: a central diffraction spike, with high spatial frequency, given by the Airy function, and a surrounding halo, with lower spatial frequency, which has a shape given by a Lorentzian function. In the case of NIFS data cubes, however, the PSF usually obtained with the use of the AO can be better described as the sum of a Gaussian and a Lorentzian function:

\begin{equation}
A \times e^{-\frac{\left(x - x_c\right)^2}{2\sigma_g^2}} + B \times \left\{\frac{1}{\left[1+\left(\frac{x - x_c}{\sigma_l^2}\right)^2\right]}\right\}.
\end{equation} 

The reason for the existence of this Gaussian component, instead of a component given by the Airy function, is probably related to the original size of NIFS spatial pixels (0.103 arcsec $\times$ 0.043\arcsec). The full width at half-maximum (FWHM) values of the Airy function in 1.25, 1.65 and 2.2 $\mu m$ (which are, approximately, the mean wavelengths of the \textit{J}, \textit{H} and \textit{K} bands, respectively,) are 0.033, 0.044 and 0.058 arcsec, respectively. We can see that all of these values are smaller than one of the dimensions of an NIFS spatial pixel (0.103 arcsec) and very close to the other (0.043 arcsec). In other words, we can say that the FWHM of the Airy function in NIFS data is always smaller than two spatial pixels (in the \textit{J} band and in part of the \textit{H} band, it is actually smaller than one spatial pixel). As the spatial sampling is not adequate to resolve the Airy profile, a Gaussian profile is usually a good description for the central spike component. In addition, the rectangular shape of NIFS spatial pixels implies that the PSF is generally non-circular. Fig.~\ref{fig18} shows graphs with the horizontal brightness profile of the image of a collapsed intermediate-wavelength interval of the data cube of GQ Lup. Superposed to these graphs are fits based on equation (13) and on a simple Gaussian function:

\begin{equation}
A \times e^{-\frac{\left(x - x_c\right)^2}{2\sigma^2}}.
\end{equation}

By analysing Fig.~\ref{fig18}, one can clearly see that the brightness profile of the image of the data cube of GQ Lup is better described by a two component fit (Gaussian + Lorentzian) than a simple Gaussian fit. This is caused by the fact that, as mentioned before, the halo around the Gaussian component, generated by the AO, has a Lorentzian shape. It is important to emphasize, however, that the relative weight between the Gaussian and the Lorentzian component in the PSF of an observation depends on the effect of the AO in the observation.

The Airy function is wavelength dependent; the NIFS spatial sampling is not. Therefore the PSFs of NIFS data cubes do not change significantly along the spectral axis. For this reason, we can apply the Richardson-Lucy deconvolution to each image of an NIFS data cube using a constant PSF, with no wavelength dependence.

Since the PSFs of AO images have usually two components, the measurement of the FWHM is not a very effective technique to evaluate the performance of the AO in the observation or even the improvement obtained with the use of a Richardson-Lucy deconvolution. A much better indicator to be used is the Strehl ratio, which is defined as

\begin{equation}
Strehl = \frac{I\left(0\right)_{PSF}}{I\left(0\right)_{Airy}},
\end{equation}
where $I(0)_{PSF}$ is the maximum intensity of the normalized PSF of the observation and $I(0)_{Airy}$ is the maximum intensity of the normalized Airy function. Since the AO is not perfect, the value of $I(0)_{PSF}$ is never equal to $I(0)_{Airy}$. In other words, we can say that, although this technique improves considerably the spatial resolution of the observation, the Strehl ratios obtained are always smaller than 1.

Due to the existence of a Gaussian and a Lorentzian component, the PSFs of NIFS data cubes are quite complex. Besides that, the Richardson-Lucy deconvolution is quite sensitive to the knowledge of the PSF. For this reason, the best results of the Richardson-Lucy deconvolution in NIFS data cubes are achieved using real PSFs, obtained directly from the data. In the case of observations of isolated stars, for example, one can use as PSF an image of an intermediate wavelength of the data cube. Observations of type 1 active galactic nuclei (AGNs) can also be used to obtain real PSFs: the broad-line region (BLR), which is the source of the broad components in permitted lines, can be regarded as a point-like source at the distance of typical AGNs; therefore, an image of a broad wing of one of the permitted lines can be used as PSF. If it is not possible to use a real image as PSF, one can also construct a synthetic PSF, based on equation (13), taking into account the fact that the PSF may not be circular. In certain cases, it may be difficult to obtain any estimate of the PSF of the observation from the analysed data cube. In these cases, one possibility is to obtain this estimate from the data cube of the standard star used in the data reduction for telluric absorption removal and flux calibration. This strategy, however, is potentially problematic because there may be differences between the effects of the AO applied to the science data cube and to the standard star data cube. There may be also differences between the seeing of these observations. Nevertheless, in many cases, these differences are of second order and the PSF obtained from the standard star data cube allows one to perform a reliable Richardson-Lucy deconvolution. In any case, one should be cautious and test if the process is really effective. Finally, for the cases in which it is absolutely impossible to obtain any estimate of the PSF (from the science data cube or from the standard star data cube), the Richardson-Lucy deconvolution should not be applied. 

One important point to be mentioned here is that the PSF of an NIFS observation may be unstable along the FOV, revealing a series of asymmetries. These irregularities (dependent on the position of the objects along the FOV) are probably caused by NIFS spatial pixels, which (as mentioned in Section 1) are not square, but rectangular and comparable to or larger than the FWHM of the Airy function. This variation of the PSF along the FOV can affect the Richardson-Lucy deconvolution. When more than one observation, with an appropriate dithering, is available, the irregularities of the PSF may be eliminated, or significantly reduced, by simply combining the data cubes in the form of a median. However, when only one observation is available and the instabilities of the PSF along the FOV are considerable, the Richardson-Lucy deconvolution should not be applied. 

Our previous experiences revealed that, in order to optimize the results obtained with the Richardson-Lucy deconvolution, a number of iterations between 6 and 10 is recommended. A Richardson-Lucy deconvolution with more than 10 iterations will not improve significantly the results obtained with 10 iterations, and will amplify the low-frequency noise in the images. A Richardson-Lucy deconvolution with less than 6 iterations will not improve sufficiently the spatial resolution of the observation to justify the effort.

Fig.~\ref{fig19} shows images of a collapsed intermediate-wavelength interval of the data cube of GQ Lup, before and after the Richardson-Lucy deconvolution, using a real PSF (corresponding to an image of the data cube in an intermediate wavelength) and 10 iterations. The measured Strehl ratios improved from Strehl = 0.08 to Strehl = 0.21. This shows that the Richardson-Lucy deconvolution resulted in a significant improvement in the spatial resolution of the data cube of GQ Lup.

One relevant topic that is worth mentioning here is the fact that, even under excellent seeing conditions, the observed Strehl ratios in NIFS data cubes are normally not higher than 0.2. We believe that these considerably low values of the Strehl ratio are not only the result of limitations in the AO, but also a consequence of the size of NIFS spatial pixels. In order to evaluate this hypothesis, we sampled Airy functions, in the \textit{J}, \textit{H} and \textit{K} bands, in spatial pixels of 0.103 arcsec $\times$ 0.043 arcsec (the same size of NIFS spatial pixels) and calculated the Strehl ratio of the resulting images. We obtained values of $\sim 0.18$, $\sim 0.30$ and $\sim 0.54$, in the \textit{J}, \textit{H} and \textit{K} bands, respectively. This result indicates that, even if the AO technique was perfect, we could not obtain Strehl ratios higher than $\sim 0.18$, $\sim 0.30$ and $\sim 0.54$ in NIFS data cubes observed in the \textit{J}, \textit{H} and \textit{K} bands, respectively.

\section{A scientific example: NGC 4151}

In order to show the benefits obtained with our data treatment procedure, we performed an analysis of the Br$\gamma$ and H$_2$ $\lambda 21218$ emission lines of the data cube of NGC 4151, in the \textit{K} band, which was treated with all the data treatment techniques described in the previous sections. These data were already published in two papers by \citet{sto09, sto10}. These papers used standard methodologies and we are reanalyzing these data with our data treatment methodology, showing new relevant scientific aspects.

\begin{figure*}
\begin{center}
  \includegraphics[scale=0.48]{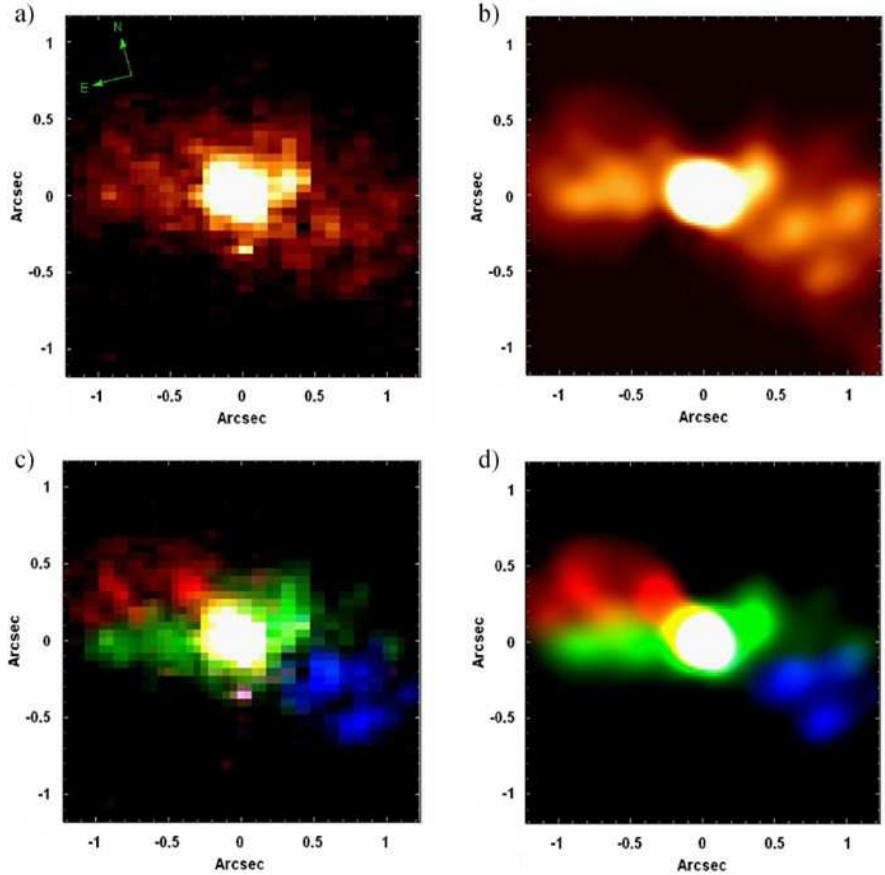}
  \caption{a) Image of the Br$\gamma$ emission line from the non-treated data cube of NGC 4151. b) Image of the Br$\gamma$ emission line from the treated data cube of NGC 4151. c) RGB composition for the image shown in a), with the colours blue, green and red representing the velocity ranges $-482$ km s$^{-1} \le V_r \le -158$ km s$^{-1}$, $-128$ km s$^{-1} \le V_r \le 167$ km s$^{-1}$ and $196$ km s$^{-1} \le V_r \le 520$ km s$^{-1}$, respectively.\label{fig20}}
\end{center}
\end{figure*}

\begin{figure*}
\begin{center}
  \includegraphics[scale=0.35]{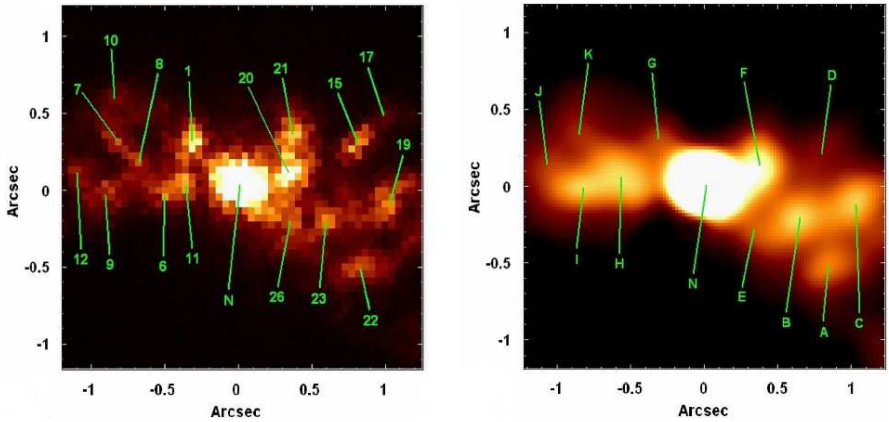}
  \caption{Left: [O III] image of the nuclear region of NGC 4151, with the same FOV of NIFS, obtained with WFPC2 of the \textit{HST}. The identification numbers of the clouds are the same ones used by \citet{hut98}. Right: image of the Br$\gamma$ emission line from the treated data cube of NGC 4151, with the detected ionized-gas clouds identified.\label{fig21}}
\end{center}
\end{figure*}

NGC 4151 is a SAB(rs)ab galaxy at a distance of 13.3 Mpc. Since it is the closest Seyfert 1 galaxy, this object is also one of the most studied AGNs. \citet{ost76} classified this galaxy as a Seyfert 1.5, as it has both broad permitted lines and strong narrow forbidden lines. NGC 4151 was in the list of objects analysed by \citet{sey43}; however, the first detailed study of the Seyfert nucleus of this galaxy was made by \citet{ulr73}, who demonstrated the existence of, at least, 4 distinct clouds in the central 300 pc.

\citet{eva93} analysed [O III] $\lambda 5007$ and H$\alpha$ + [N II] $\lambda \lambda 6548,6583$ images of the nuclear region of NGC 4151, obtained with the \textit{Hubble Space Telescope} (\textit{HST}). The authors concluded that the H$\alpha$ + [N II] $\lambda \lambda 6548,6583$ emission comes from an unresolved nuclear point-like source, probably associated with the BLR of the AGN. On the other hand, the [O III] $\lambda 5007$ emission, associated with the narrow-line region (NLR), comes from clouds distributed under the form of a bicone. The authors also verified that the bicone has a projected opening angle of $75\degr \pm 5\degr$ and extends along a position angle of $60\degr \pm 5\degr$.

By analysing images and spectra from the \textit{HST}, \citet{hut98} determined velocities of individual clouds in the NLR of NGC 4151 and concluded that the model of outflows along the bicone is the most appropriate to describe the observed behaviour. The authors also concluded that the edge of the ionization cone must be close to the line of sight. \citet{cre00} and \citet{das05} also analysed data from the \textit{HST} of the nuclear region of NGC 4151 and elaborated models of biconical outflows. They concluded that the main aspects of the ionized-gas kinematics in the NLR of this galaxy could be reproduced by these models.

\citet{kno96} analysed spectroscopic data of NGC 4151 in the near-infrared (1.24 - 1.30 $\mu m$) and identified individual velocity components for the Pa$\beta$, [Fe II] $\lambda 12567$, and [SI X] $\lambda 12524$ emission lines, which is consistent with the presence of individual clouds. 

\begin{table*}
\begin{center}
\caption{Comparison between the ionized-gas clouds in the nuclear region of NGC 4151 identified using NIFS data and \textit{HST} data (Hutchings et al. 1998, using \textit{HST} data).\label{tbl1}}
\begin{tabular}{cccc}
\hline
Identification of the cloud      & Identification of the cloud   &  Radial velocity obtained             &  Radial velocity obtained by \\
in the image of Br$\gamma$  & in the [O III] image, made   &  using NIFS data (km s$^{-1}$)  & \citet{hut98}, using   \\
                                               & by \citet{hut98}      &                                                     & \textit{HST} data (km s$^{-1}$)  \\
\hline
A  &  $22$             &  $-238 \pm 2$   &  $-260 \pm 40$\\
B  &  $23$             &  $-222 \pm 2$   &  $-213 \pm 3$\\
C  &  $19$             &  $-173 \pm 5$   &  $-126 \pm 32$\\
D  &  $15+17$  &  $-18 \pm 7$     &  $-210 \pm 12$ (cloud $15$)\\
    &                   &                          &  $-210 \pm 11$ (cloud $17$)\\
E  &  $26$         &  $-200 \pm 9$   &  $-340 \pm 20$\\
F  &  $20$         &  $17 \pm 5$      &  $37 \pm 8$\\
G  &  $1$          &  $222 \pm 17$  &  $223 \pm 9$\\
H  &  $6+11$    &  $36 \pm 2$      &  $12 \pm 21$ (cloud $6$)\\
    &                   &                          &  $20 \pm 18$ (cloud $11$)\\
I  &  $9$           &  $24 \pm 2$      &  $60 \pm 18$\\
J  &  $12$         &  $122 \pm 11$  &  $100 \pm 85$\\
K  & $7+8$       &  $213 \pm 8$    &  $235 \pm 70$ (cloud $7$)\\
    &                   &                          &  $311 \pm 47$ (cloud$8$)\\
\hline
\end{tabular}
\end{center}
\end{table*}

\citet{sto09} verified that the intensity distributions of the H, He, [S III], [P II], [Fe II] and [S III] emission lines have a morphology of a bicone, as expected from previous studies. The intensity distribution of H$_2$, on the other hand, is completely different, as almost no emission of H$_2$ was detected along the bicone, possibly due to the destruction of the H$_2$ molecules in this area. The authors proposed that the H$_2$ emission is probably a tracer of a large molecular gas reservoir, which feeds the AGN. \citet{sto10} identified three kinematical components for the ionized gas: an extended emission, with a systemic velocity, in a circular region around the nucleus; an outflow component along the bicone; a component due to the interaction of the radio jet with the disc of the galaxy.

\begin{figure}
\begin{center}
  \includegraphics[scale=0.23]{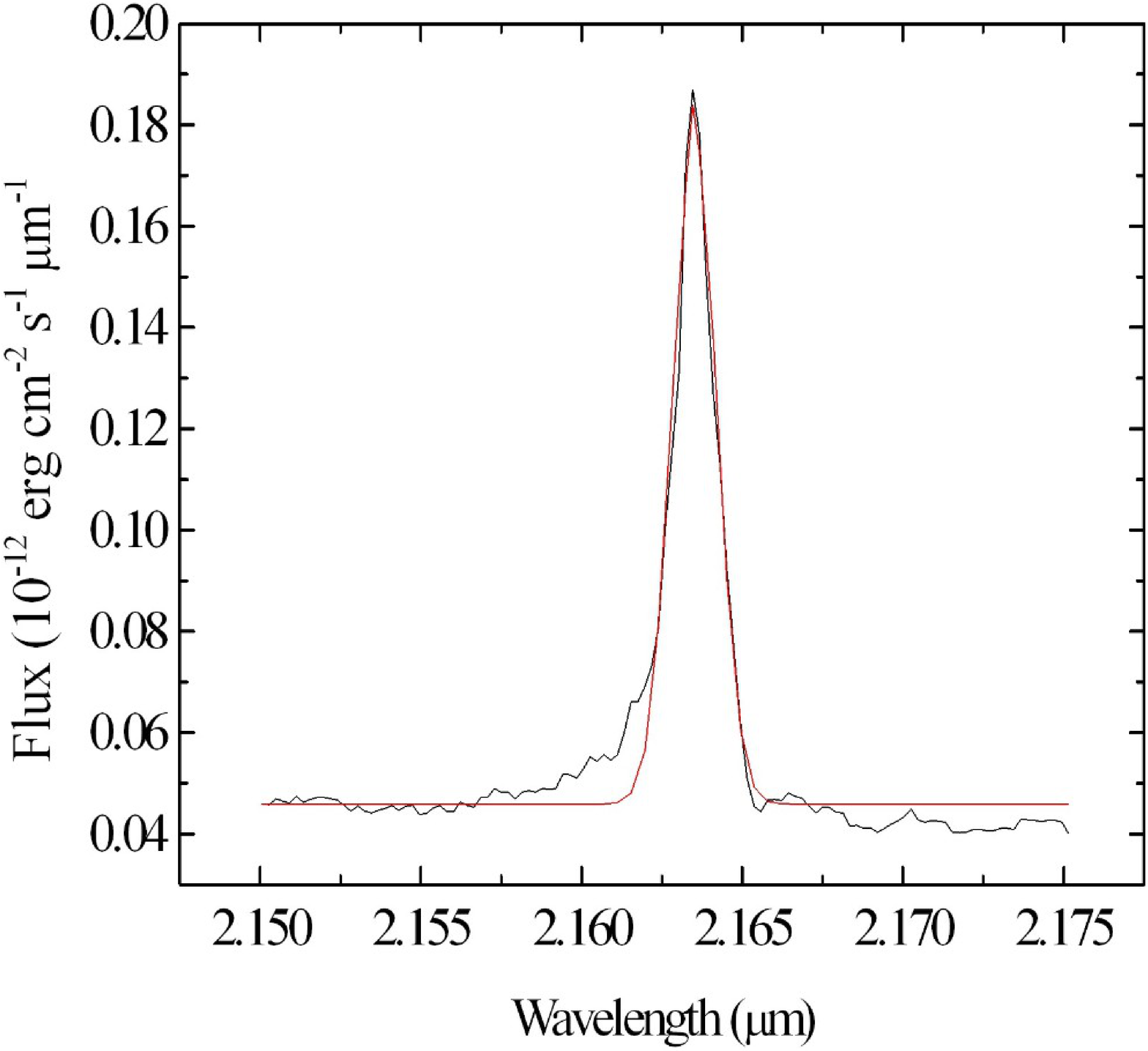}
  \caption{Gaussian fit (in red) applied to the Br$\gamma$ emission line of the spectrum of cloud A from the treated data cube of NGC 4151.\label{fig22}}
\end{center}
\end{figure}

In order to evaluate the effects of our data treatment procedure in the analysis of the data cube of NGC 4151, first of all, we constructed images of the Br$\gamma$ emission line (with the adjacent spectral continuum subtracted) from the data cubes of NGC 4151 with and without the use of our data treatment procedure. We also elaborated RGB compositions (based on the radial velocity values) for the images obtained. The results are shown in Fig.~\ref{fig20}. All the images reveal a biconical morphology for the NLR of this galaxy, as observed by many previous studies \citep{eva93, hut98,cre00, das05, sto09, sto10}. However, the visualization of this morphology is more difficult in the Br$\gamma$ image from the non-treated data cube. Our data treatment procedure not only improved the visualization of the bicone, but also made it easier to detect even some individual ionized-gas clouds, which are much harder to identify (some of them are almost undetectable) in the non-treated data cube.

We retrieved an [O III] image of the nuclear region of NGC 4151, obtained with Wide Field Planetary Camera 2 (WFPC2) of the \textit{HST} (this image was analysed by Hutchings et al. 1998). Fig.~\ref{fig21} shows a comparison between this [O III] image and the Br$\gamma$ image obtained from the treated data cube of NGC 4151. We can see that, although our spatial resolution is not as high as the spatial resolution obtained with the \textit{HST} (as expected), the improvement provided by our data treatment procedure is so high that many of the ionized-gas clouds visible in the [O III] image are also very clearly detected in the Br$\gamma$ image from the treated data cube. Some pairs of clouds ($15+17$, $6+11$, and $7+8$), however, remained blended in the Br$\gamma$ image. Table~\ref{tbl1} shows a comparison between the ionized-gas clouds identified by \citet{hut98} and the ones identified in the Br$\gamma$ image from the treated data cube.

\begin{figure*}
\begin{center}
  \includegraphics[scale=0.48]{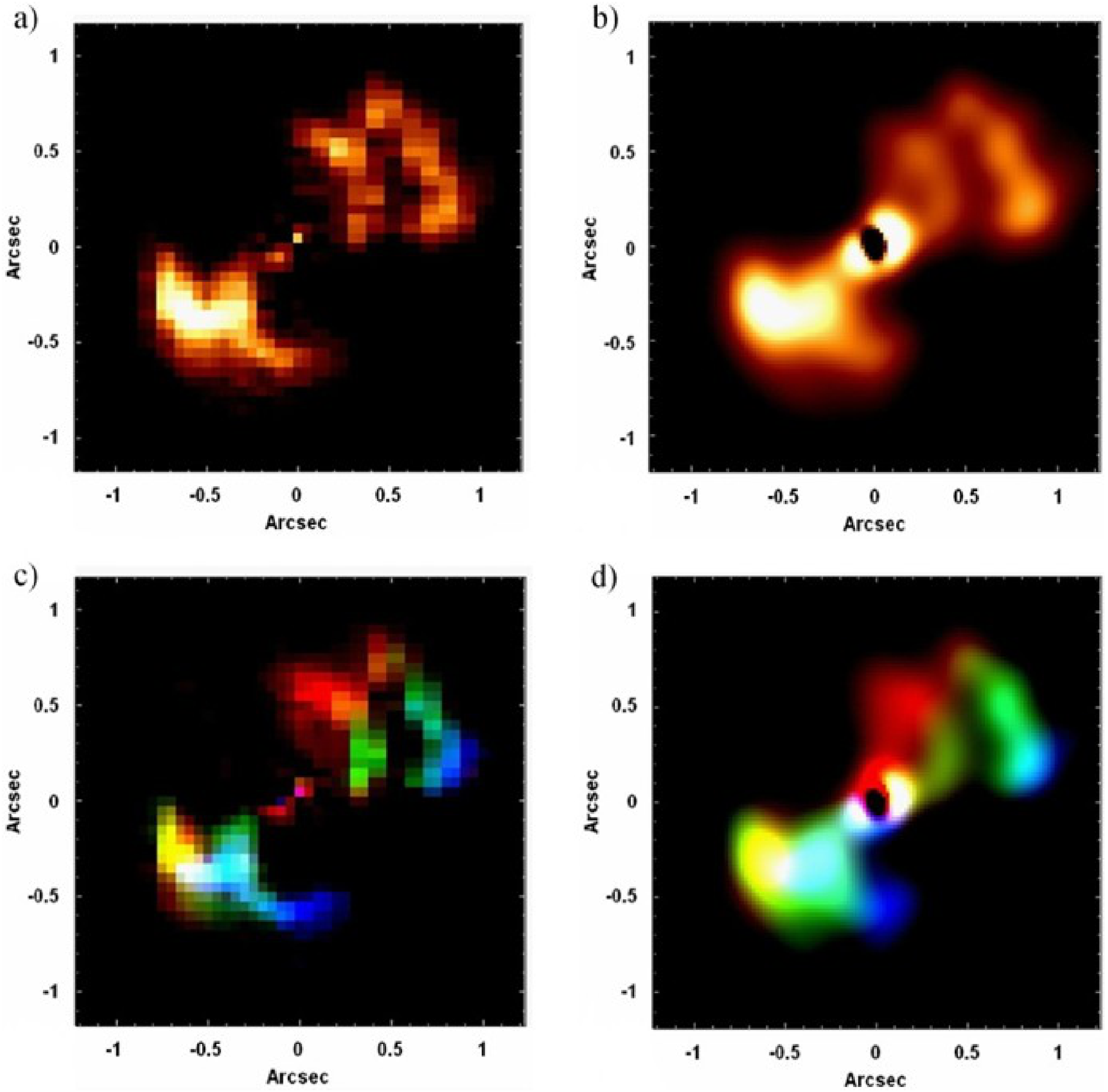}
  \caption{a) Image of the H$_2$ $\lambda 21218$ emission line from the non-treated data cube of NGC 4151. b) Image of the H$_2$ $\lambda 21218$ emission line from the treated data cube of NGC 4151. c) RGB composition for the image shown in a), with the colours blue, green and red representing the velocity ranges $-94$ km s$^{-1} \le V_r \le 26$ km s$^{-1}$, $56$ km s$^{-1} \le V_r \le 86$ km s$^{-1}$ and $116$ km s$^{-1} \le V_r \le 236$ km s$^{-1}$, respectively. d) RGB composition for the image shown in b), with the colours blue, green and red representing the velocity ranges $-94$ km s$^{-1} \le V_r \le 26$ km s$^{-1}$, $56$ km s$^{-1} \le V_r \le 86$ km s$^{-1}$ and $116$ km s$^{-1} \le V_r \le 236$ km s$^{-1}$, respectively.\label{fig23}}
\end{center}
\end{figure*}

We extracted spectra from each one of the clouds we could identify in the Br$\gamma$ image from the treated data cube. After that, we shifted each spectrum to the rest frame, using \textit{z} = 0.003319 (NASA Extragalactic Database - NED), and also applied a correction due to the heliocentric velocity. These two steps were applied, in the IRAF environment, using the tasks `\textit{dopcor}' (from the `NOAO' package) and `\textit{rvcorret}' (from the `ASTUTIL' package), respectively. We, then, determined the radial velocity of the ionized gas for each spectrum by fitting Gaussian functions to the Br$\gamma$ emission line. A similar work had already been done by Hutchings et al. (1998); however, they determined the radial velocities of the clouds using an association of Space Telescope Imaging Spectrograph (STIS) spectra and images obtained with WFPC2. This method involving these two instruments of the \textit{HST} resulted in high uncertainties for the radial velocities. Our procedure reduced substantially the uncertainties obtained for the radial velocities. Fig.~\ref{fig22} shows an example of the Gaussian fit applied to the Br$\gamma$ emission line of the spectrum of one of the ionized-gas clouds (cloud A) detected in the treated data cube of NGC 4151. The fit provides, with a considerable precision, the wavelength corresponding to the peak of the emission line and, therefore, the radial velocity. All of the Gaussian fits applied to the spectra of the other clouds were also able to provide, with similar precisions, reliable values for the radial velocities of the ionized gas.

Table~\ref{tbl1} shows a comparison between the radial velocities obtained by us and by \citet{hut98}. It is easy to see that the velocity values obtained for each cloud in the two works are all compatible, at $3\sigma$ level, with the exception of clouds D and E. There were no irregularities or difficulties in the Gaussian fits of the Br$\gamma$ emission line in the spectra of these two clouds that could explain these differences. One hypothesis is that the uncertainties obtained by \citet{hut98} may be underestimated for these two clouds.

Since we have a limited FOV, we cannot evaluate the kinematical behaviour of the ionized gas at greater distances from the AGN. However, Fig.~\ref{fig21} and Table~\ref{tbl1} show that the two clouds with the highest projected distances from the AGN (clouds C and J) have radial velocities significantly lower than many of the clouds at shorter projected distances. These results suggest that the clouds are accelerated near the AGN and then are decelerated at farther distances, probably due to the interaction with the surrounding medium. Such behaviour is compatible with the models of biconical outflows proposed by \citet{cre00} and \citet{das05}.

We also constructed images of the H$_2$ $\lambda 21218$ emission line (with the adjacent spectral continuum subtracted) from the data cubes of NGC 4151 with and without the use of our data treatment procedure. These images are shown in Fig.~\ref{fig23}, together with RGB compositions based on the radial velocity values. All the images reveal the existence of structures similar to `arcs' along a direction perpendicular to the ionization cone of the galaxy. This behaviour is probably due to the destruction of the H$_2$ by the ionizing flux along the cone. Similar conclusions were also obtained by \citet{sto09}. The visualization of these arcs was significantly improved by our data treatment. Substructures similar to `knots' can be seen in the image from the treated data cube, but are much harder to detect in the image from the non-treated data cube. One of the most important improvements obtained with the data treatment procedure is the revelation of structures connecting the arcs and the inner region, where the AGN is located. The most well defined of these structures appears in green color in the RGB image in Fig.~\ref{fig23}. Although a subtle connection between the arcs and the inner region is visible in the image from the non-treated data cube, the visualization of this connection is severely improved in the image from the treated data cube. Our interpretation is that these connecting structures may represent flows of material feeding the central AGN. This scenario is compatible with the hypothesis proposed by \citet{sto09}, in which the H$_2$ emission is a tracer of a large molecular gas reservoir, which feeds the AGN.

\section{Summary and conclusions}

\begin{figure*}
\begin{center}
  \includegraphics[scale=0.46]{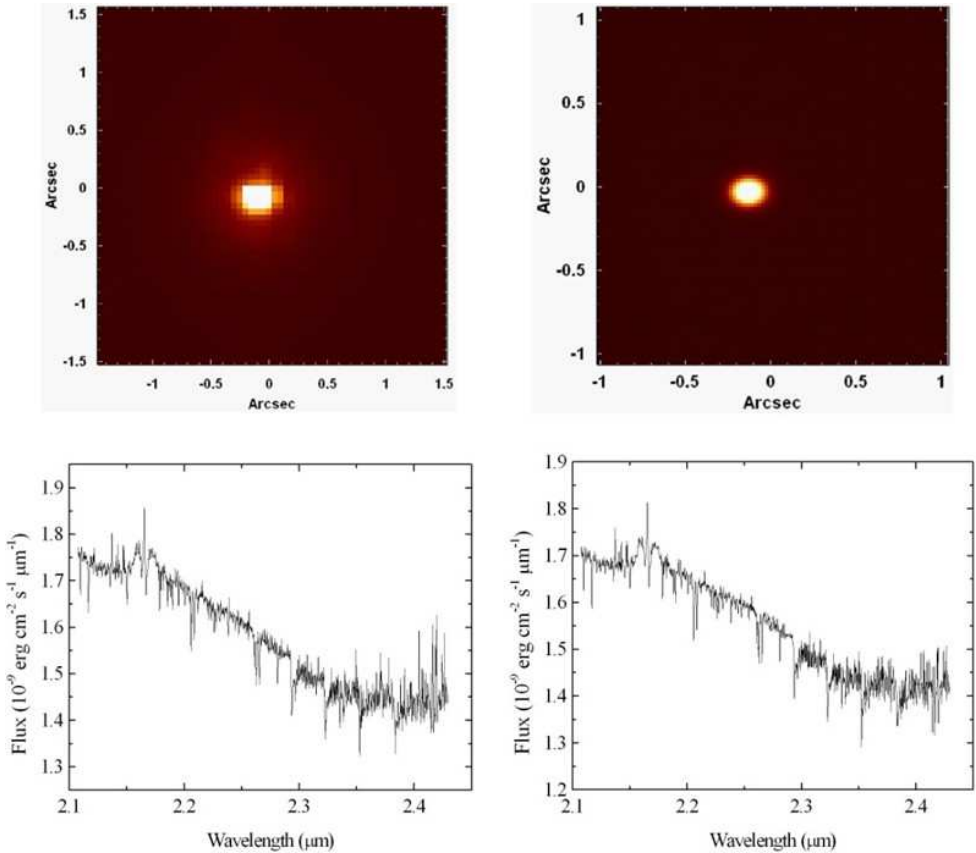}
  \caption{Left: image of one of the nine data cubes of GQ Lup, collapsed along the spectral axis, obtained after the data reduction. Right: image of the final data cube of GQ Lup, also collapsed along the spectral axis, obtained after the entire data treatment. The corresponding total spectra, extracted from each one of the previous data cubes, are also shown.\label{fig24}}
\end{center}
\end{figure*}

In this paper, we presented a detailed procedure for treating NIFS data cubes, after the data reduction. This treatment includes the following steps: correction of the DAR, spatial re-sampling, Butterworth spatial filtering, instrumental fingerprint removal and Richardson-Lucy deconvolution. 

We verified that the total spatial displacement of structures along the spectral axis of NIFS data cubes, considering the \textit{J}, \textit{H} and \textit{K} bands together is larger than the typical NIFS spatial resolution (0.1 arcsec) in certain cases. Therefore, this effect may be significant for studies involving the entire spectral continuum of the \textit{J}, \textit{H} and \textit{K} bands and should be corrected. One example of such a study is a spectral synthesis. Studies involving spectral lines may also be affected by this effect if one wishes to compare images of lines located in different spectral bands. On the other hand, the spatial displacements of structures along the spectral axis of NIFS data cubes in each one of these three spectral bands are usually smaller than the typical NIFS spatial resolution. Therefore, studies involving only one of the spectral bands mentioned above may not require a correction of this effect. The possibility of these spatial displacements being significant in NIFS data cubes may be unexpected, as a correction of the DAR effect is usually ignored in the infrared. However, this can be explained by the use of AO in NIFS observations, resulting in very high spatial resolutions. Therefore, although the DAR is small in the infrared, the high spatial resolutions of NIFS observations provided by the AO make this effect significant in certain cases. The AO-corrected PSFs obtained with the next generation of extremely large telescopes will result in a larger impact of the DAR effect on data cubes.

The theoretical curves obtained assuming a plane-parallel atmosphere or taking into account the curvature of the atmosphere do not reproduce the exact behaviour of the spatial displacements along the spectral axis of NIFS data cubes. Our tests revealed that this discrepancy cannot be attributed to the spatial re-sampling applied during the data reduction. The exact cause of this problem remains unknown. One possibility is that this is an instrumental effect. Since the theoretical curves do not reproduce properly the spatial displacements along the spectral axis, the practical approach is the most precise for removing this effect from NIFS data cubes. However, since the discrepancies between the theoretical curves and the observed spatial displacements are usually smaller than the typical NIFS spatial resolution, we conclude that the theoretical approach may also be used to remove, with a good precision, the spatial displacements of structures along the spectral axis of NIFS data cubes.

The spatial re-sampling, only, does not improve significantly the appearance of the images in the data cubes because, in order to conserve the surface flux, the re-sampled images end up with groups of adjacent spaxels with equal or similar values. In order to solve this problem, after the spatial re-sampling of an image, we also apply an interpolation in the values of the spaxels in each one of the lines and, after that, in each one of the columns of the image. There are many interpolation methods that can be used and our tests revealed that the most appropriate of them are: lsquadratic, quadratic and spline. All of these methods provide similar results and, therefore, may be used in the data treatment procedure. The spatial re-sampling followed by an interpolation results in a considerable improvement of the appearance of the images in the data cubes. One side effect of applying this spatial re-sampling is a degradation of the S/N ratio of the spectra in the data cubes. However, our previous experiences revealed that this degradation is usually lower than 1.0 per cent. Considering these very low degradations and the significant improvements obtained in the visualization of the contours of the structures, we conclude that the spatial re-sampling is a procedure that is worth applying to NIFS data cubes. Another side effect of the spatial re-sampling is that it introduces high spatial-frequency components in the images. However, these components are almost entirely removed by the Butterworth spatial filtering.

The Butterworth spatial filtering, with order equal to 2 and an appropriate cut-off frequency, removes a considerable amount of high-frequency noise (in the form of thin vertical and horizontal stripes) from the images of NIFS data cubes, without introducing significant artefacts or affecting the PSFs of the observations. A filtering applied with order higher than 2 introduces concentric rings around point-like sources in the images. These rings are only clearly detectable in the $W_0$ data cube, obtained with a wavelet decomposition. The $W_0$ data cube can also be used to determine an appropriate value for the cut-off frequency. Our tests revealed that the most appropriate Butterworth spatial filtering for NIFS data cubes uses a filter corresponding to the product of an elliptical filter by a rectangular filter and cut-off frequencies ranging from 0.19 Ny to more than 0.40 Ny. 

We observed that many of the NIFS data cubes show instrumental fingerprints in the form of vertical stripes in the images. These fingerprints also have a very specific spectral signature. We use the PCA Tomography technique to isolate and remove these instrumental fingerprints.

The Richardson-Lucy deconvolution results in a considerable increase of the Strehl ratio of NIFS data cubes, indicating a significant improvement of the spatial resolution. The PSFs to be used for this procedure are composed of a Gaussian component and a Lorentzian component. The best results are achieved using real PSFs, obtained directly from the data. If it is not possible to use a real image as PSF, one can also construct a synthetic PSF. However, when it is not possible to obtain a reliable estimate of the PSF of the observation, the Richardson-Lucy deconvolution should not be applied. Our experience suggests that a number of iterations between 6 and 10 is recommended, in order to obtain optimized results.

The analysis of the data cubes of NGC 4151 revealed very clearly the improvements provided by our data treatment procedure. An image of the Br$\gamma$ emission line from the treated data cube shows a biconical morphology of the NLR of this galaxy, as already observed in many previous studies \citep{eva93, hut98,cre00, das05, sto09, sto10}. The visualization of this morphology is more difficult in the Br$\gamma$ image from the non-treated data cube. However, our data treatment not only improved the visualization of the bicone, but also made it easier to detect even some individual ionized-gas clouds, which are much harder to identify (some of them are almost undetectable) in the non-treated data cube. The resolution of the Br$\gamma$ image from the treated data cube was so improved that we could even establish a comparison with an [O III] image of the nuclear region of NGC 4151, obtained with WFPC2 of the \textit{HST}. Although our spatial resolution, as expected, is not as high as the spatial resolution obtained with the \textit{HST}, many of the ionized-gas clouds visible in the [O III] image are also very clearly detected in the Br$\gamma$ image from the treated data cube. We determined the radial velocity of each ionized-gas cloud we could identify in the Br$\gamma$ image. A similar work had already been done by Hutchings et al. (1998); however, their methodology (using data from STIS and WFPC2 of the \textit{HST}) resulted in high uncertainties for the radial velocities. Our procedure reduced substantially the uncertainties obtained for the radial velocities. Since we have a limited FOV, we cannot evaluate the kinematical behaviour of the ionized gas at greater distances from the AGN. However, our results suggest that the clouds are accelerated near the AGN and then are decelerated at farther distances, probably due to the interaction with the surrounding medium. Such behaviour is compatible with the models of biconical outflows proposed by \citet{cre00} and \citet{das05}.

The H$_2$ $\lambda 21218$ image from the treated data cube revealed the existence of structures similar to `arcs' along a direction perpendicular to the ionization cone. This behaviour is probably due to the destruction of the H$_2$ by the ionizing flux along the cone. Similar conclusions were also obtained by \citet{sto09}. The visualization of these arcs was significantly improved by our data treatment. However, the H$_2$ image from the treated data cube also revealed the existence of structures connecting the arcs and the inner region, where the AGN is located. The visualization of such connection was much more difficult in the image from the non-treated data cube. Our interpretation is that these connecting structures may represent flows of material feeding the central AGN. This scenario is compatible with the hypothesis proposed by \citet{sto09}, in which the H$_2$ emission is a tracer of a large molecular gas reservoir, which feeds the AGN.

One important final question is: what is the effect of our data treatment procedure on the spectra of the data cubes? Fig.~\ref{fig24} shows a comparison between the image of one of the nine data cubes of GQ Lup in the \textit{K} band, collapsed along the spectral axis, obtained after the data reduction, and the image of the final data cube of GQ Lup in the \textit{K} band, also collapsed along the spectral axis, obtained after the entire data treatment. In the same figure, we have the total spectra of these two data cubes.

The improvement in the images of the data cube achieved with the treatment procedure is very clear in Fig.~\ref{fig24}. The clearer contours of the structures obtained with the spatial re-sampling, the high spatial-frequency noise removed with the Butterworth spatial filtering, and the improvement of the spatial resolution obtained with the Richardson-Lucy deconvolution resulted in images with a considerably higher quality. On the other hand, the improvement in the spectra of the data cube is much more subtle. We can see, for example, that some of the high spectral-frequency structures in the red region of the spectrum were removed by the treatment procedure, but the S/N ratio did not change significantly. This is expected, as the methodologies we describe in this work are focused on the images and not on the spectra of the data cube. There are methodologies that can be used to reduce the spectral noise in certain cases, like a Butterworth spectral filtering, for example; however, we will not give more details about these procedures in this work.

Considering the significant improvement of the images of NIFS data cubes we obtained by applying our treatment procedure, we believe that this methodology may result in more reliable analysis of data obtained with this instrument and, therefore, may be of interest of the astronomical community. The scripts of all procedures described in this paper can be obtained at \textit{http://www.astro.iag.usp.br/$\sim$PCAtomography}.

\section*{Acknowledgements}

This work is based on observations obtained at the Gemini Observatory, which is operated by the Association of Universities for Research in Astronomy, Inc., under a cooperative agreement with the NSF on behalf of the Gemini partnership: the National Science Foundation (United States), the Science and Technology Facilities Council (United Kingdom), the National Research Council (Canada), CONICYT (Chile), the Australian Research Council (Australia), Minist\'erio da Ci\^encia, Tecnologia e Inova\c{c}\~ao (Brazil) and Ministerio de Ciencia, Tecnolog\'ia e Innovaci\'on Productiva  (Argentina). We would like to thank FAPESP for support under grants 2012/02268-8 (RBM) and 2012/21350-7 (TVR).

\end{document}